\documentclass[12pt,english, journal,draftclsnofoot,onecolumn]{IEEEtran}
\usepackage[T1]{fontenc}
\usepackage[latin9]{inputenc}
\usepackage{float}
\usepackage{mathrsfs}
\usepackage{amsthm}
\usepackage{amsmath}
\usepackage{amssymb}
\usepackage{stackrel}
\usepackage{graphicx}
\usepackage{setspace}
\usepackage{esint}
\makeatletter

\floatstyle{ruled}
\newfloat{algorithm}{tbp}{loa}
\providecommand{\algorithmname}{Algorithm}
\floatname{algorithm}{\protect\algorithmname}

\floatstyle{ruled}
\newfloat{algorithm}{tbp}{loa}
\providecommand{\algorithmname}{Algorithm}
\floatname{algorithm}{\protect\algorithmname}

\theoremstyle{plain}
\newtheorem{thm}{\protect\theoremname}

\usepackage{colortbl}
\usepackage{cite}
\usepackage{nomencl}
\usepackage{flushend}
\usepackage{stfloats}
\usepackage[caption=false]{subfig}

\@ifundefined{showcaptionsetup}{}{%
\PassOptionsToPackage{caption=false}{subfig}}
\usepackage{subfig}
\makeatother

\usepackage{babel}
\addto\captionsamerican{\renewcommand{\algorithmname}{Algorithm}}
\addto\captionsamerican{\renewcommand{\theoremname}{Theorem}}
\addto\captionsenglish{\renewcommand{\theoremname}{Theorem}}
\providecommand{\theoremname}{Theorem}

\begin{document}

\title{User Clustering for STAR-RIS Assisted Full-Duplex NOMA Communication
Systems}

\author{Abdelhamid Salem,\textit{\normalsize{} Member, IEEE}{\normalsize{},}
Kai-Kit Wong, \textit{\normalsize{}Fellow, IEEE}, Chan-Byoung Chae,
\textit{\normalsize{}Fellow, IEEE} and Yangyang Zhang\\
\thanks{Abdelhamid Salem is with the department of Electronic and Electrical
Engineering, University College London, London, UK, (emails: a.salem@ucl.ac.uk).
Abdelhamid Salem is also affiliated with department of Electronic
and Electrical Engineering, Benghazi University, Benghazi, Libya.

K.-K. Wong is with the department of Electronic and Electrical Engineering,
University College London, London, UK, (email: kai-kit.wong@ucl.ac.uk).
Kai-Kit Wong is also affiliated with Yonsei Frontier Lab., Yonsei
University, Seoul, 03722 Korea. 

C.-B. Chae is with the School of Integrated Technology, Yonsei University,
Seoul 03722, South Korea, (email: cbchae@yonsei.ac.kr). Y. Zhang is
with Kuang-Chi Science Ltd., Hong Kong SAR, China.

The work is supported by the Engineering and Physical Sciences Research
Council (EPSRC) under grant EP/V052942/1 and by IITP grant funded
by MSIT (2022-0-00704). For the purpose of open access, the authors
will apply a Creative Commons Attribution (CCBY) license to any Author
Accepted Manuscript version arising.%
} }
\maketitle
\begin{abstract}
In contrast to conventional reconfigurable intelligent surface (RIS),
simultaneous transmitting and reflecting reconfigurable intelligent
surface (STAR-RIS) has been proposed recently to enlarge the serving
area from $180^{o}$ to $360^{o}$ coverage. This work considers the
performance of a STAR-RIS aided full-duplex (FD) non-orthogonal multiple
access (NOMA) communication systems. The STAR-RIS is implemented at
the cell-edge to assist the cell-edge users, while the cell-center
users can communicate directly with a FD base station (BS). We first
introduce new user clustering schemes for the downlink and uplink
transmissions. Then, based on the proposed transmission schemes closed-form
expressions of the ergodic rates in the downlink and uplink modes
are derived taking into account the system impairments caused by the
self interference at the FD-BS and the imperfect successive interference
cancellation (SIC). Moreover, an optimization problem to maximize
the total sum-rate is formulated and solved by optimizing the amplitudes
and the phase-shifts of the STAR-RIS elements and allocating the transmit
power efficiently. The performance of the proposed user clustering
schemes and the optimal STAR-RIS design are investigated through numerical
results.\end{abstract}


\section{Introduction}

Reconfigurable intelligent surface (RIS) has been envisioned as a
promising technology for sixth-generation (6G) wireless communication
networks \cite{ref1,Reef1,RISme,https://doi.org/10.48550/arxiv.2301.00276}.
RIS is capable of manipulating the propagation of electromagnetic
waves. More specifically, the conventional RIS consists of controllable
reflecting elements that can manipulate the phase shifts of the impinging
signals to improve the quality of the received signals \cite{Ref22,Ref33,Ref44,Ref55}.
However, the single-faced structure of the conventional RIS limits
its service to only half-space, and the base station (BS) and the
users should be situated on the same RIS side. To tackle this shortcoming,
very recently simultaneous transmitting and reflecting RIS (STAR-RIS)
has been proposed and investigated in the literature \cite{mag1,refone,ref9a,Ref3,Ref5,Ref6,Ref7,Ref8,Ref9,ref9b}.
STAR-RIS can simultaneously transmit and reflect the impinging signals
to the users located on both sides of the surface, and thus it can
extend the serving area from $180^{o}$ to $360^{o}$ coverage. The
concept of STAR-RIS, its potential benefits, and the key differences
between conventional RIS and STAR-RIS have been discussed in \cite{mag1}
and \cite{refone}. The authors in \cite{mag1} and \cite{refone}
proposed three practical STAR-RIS operating protocols, namely, mode
switching (MS), time switching (TS), and energy splitting (ES). In
MS protocol the STAR-RIS elements are divided into a transmission
group and a reflecting group, and in TS protocol the STAR-RIS elements
periodically switch between the transmission and reflection modes.
Whilst, ES protocol splits the energy of the incident signal on each
element into two portions for transmitting and reflecting. In \cite{ref9a}
a STAR-RIS-assisted two-user downlink (DL) communication systems has
been considered for orthogonal multiple access (OMA) and non-orthogonal
multiple access (NOMA) techniques. The performance of STAR-RIS empowered
NOMA systems to support low-latency communications has been studied
in \cite{Ref3}. Analytical expressions to evaluate the ergodic rate
and the coverage probability of a STAR-RIS assisted NOMA multi-cell
networks were provided in \cite{Ref5}. In \cite{Ref6} a power minimization
problem for STAR-RIS-aided uplink (UL) NOMA systems has been studied.
The energy efficiency of a STAR-RIS aided multiple-input and multiple-output
(MIMO)-NOMA systems with a user-pairing scheme has been considered
in \cite{Ref7}. Approximate analytical expressions of the ergodic
rate for a STAR-RIS-aided DL NOMA systems with statistical channel
state information (CSI) have been derived in \cite{Ref8}. In \cite{Ref9}
the achievable sum rate of a STAR-RIS assisted NOMA systems has been
maximized by optimizing the decoding order and the transmission-reflection
beamforming. In addition, analytical expressions of the ergodic rate
and the outage probability for a pair of users in a STAR-RIS assisted
NOMA systems over Rician fading channels have been derived in \cite{ref9b}.

Moreover, full-duplex (FD) technique allows wireless devices to simultaneously
transmit and receive messages in the same time and frequency resources
\cite{Ref10,Ref11}. Consequently, FD communication systems can offer
more flexible usage of the spectrum. Interestingly, the implementation
of RIS in FD communication systems has been proposed and studied recently
in the literature \cite{ref14,ref15,Ref12,Ref17,part1}. In \cite{ref14}
a RIS has been deployed to cover the dead zones of FD cellular communication
systems, the results in this work showed that the spectral efficiency
can be doubled compared to the conventional half duplex (HD) systems.
In \cite{ref15} a joint beamforming design for a RIS-aided multiple-antennas
FD communication systems has been studied, where the active beamforming
at the transmitter and the passive beamforming at the RIS were designed.
In \cite{Ref12} the deployment design for a RIS-assisted FD communication
system has been considered, in which an FD transmitter communicates
with an UL user and a DL user over the same time and frequency resources
through a RIS. A multiple users FD communication system was studied
in \cite{Ref17} where a dedicated RIS was assigned to each user in
the system. In our previous work \cite{part1}, we investigated the
performance of a STAR-RIS-assisted FD-NOMA pairing communication systems.

Furthermore, it has been shown in the literature that, the detection
complexity of NOMA increases as number of the users increases due
to the increase in the SIC layers. To overcome this issue, a user
clustering scheme has been introduced and investigated \cite{clus1,clus3,clus4,clus5,cluserNEW}.
For instance, in \cite{clus1} a comprehensive user clustering strategy
for DL NOMA systems has been introduced, in which the users grouped
in a cluster can receive their signals simultaneously using NOMA technique,
and time-division multiple access (TDMA) scheme has been implemented
among the different clusters. A novel user clustering scheme for NOMA
assisted multi-cell massive MIMO systems was proposed in \cite{clus3}.
Efficient user clustering schemes for UL and DL NOMA transmissions
have been designed in \cite{clus4}. In \cite{clus5} a user scheduling
problem for DL NOMA systems has been studied, in which the BS allocates
the spectrum and power resources to a set of users by taking into
account the users' fairness. In addition, a user clustering scheme
for UL multiple-input single-output (MISO)-NOMA systems has been considered
in \cite{cluserNEW}.

Accordingly, this paper considers a STAR-RIS assisted multiple-user
FD communication system, where the FD-BS employs NOMA technique to
serve the DL and UL users. The cell-edge users communicate with the
FD-BS via a passive STAR-RIS, while the cell-center users can communicate
directly with the FD-BS. The BS and the STAR-RIS are assumed to know
only the statistical channel state information (CSI) and the users'
distances. We first present new simple and efficient user-clustering
schemes for DL and UL NOMA transmissions. Then, according to the proposed
transmission schemes, the ergodic rates for the DL and UL modes are
analyzed. In addition, the total sum rates are maximized by optimizing
the STAR-RIS elements and the power transmissions. For clarity, the
main contributions are listed as follows: 

1) New user clustering schemes for STAR-RIS-aided FD NOMA communication
systems are presented and discussed.

2) New closed-form analytical expressions for the ergodic rates of
DL and UL users under Rician fading channels are derived. This channel
model is more general but also challenging and hard to analyze. The
impact of the imperfect SIC and the self interference at the FD-BS
are also taken into account in the analysis. 

3) The optimal STAR-RIS phase shifts and amplitudes that maximize
the ergodic sum-rates are obtained. In addition, sub-optimal designs
of the STAR-RIS phase shifts and amplitudes are also provided. 

4) An efficient power allocation scheme that enhances the ergodic
sum-rates is considered. 

5) Monte-Carlo simulations are executed to investigate the performance
of the user clustering schemes and the optimal system design.

The results in this work show that increasing the transmit signal
to noise ratio (SNR) always improves the achievable rates, and the
performance of the cell-edge users can be enhanced by increasing number
of the STAR-RIS elements. Furthermore, NOMA user clustering scheme
can achieve a higher sum rate than NOMA pairing scheme in the DL mode
with perfect and imperfect SIC, when $0\%$ and $10\%$ of the power
remains as interference. Whilst, in the UL mode NOMA user clustering
scheme outperforms NOMA paring scheme in the perfect SIC case, $0\%$,
and in the low SNR regime when $10\%$ of the power remains as interference. 

Next, Section \ref{sec:System-Model} presents the system model. Section
\ref{sec:Cell-edge-users-added} presents DL and UL NOMA user clustering
schemes, and Section \ref{sec:Ergodic-Rate-Analysis} analyzes the
ergodic rates of the DL and UL users. In Section \ref{sec:System-Design}
we consider the optimal system design. Section \ref{sec:Numerical-Results}
depicts our simulation and numerical results. The main conclusions
are summarized in Section \ref{sec:CONCLUSIONs}.

\section{System Model\label{sec:System-Model}}

\begin{figure}
\noindent \begin{centering}
\includegraphics[bb=80bp 10bp 780bp 500bp,clip,scale=0.35]{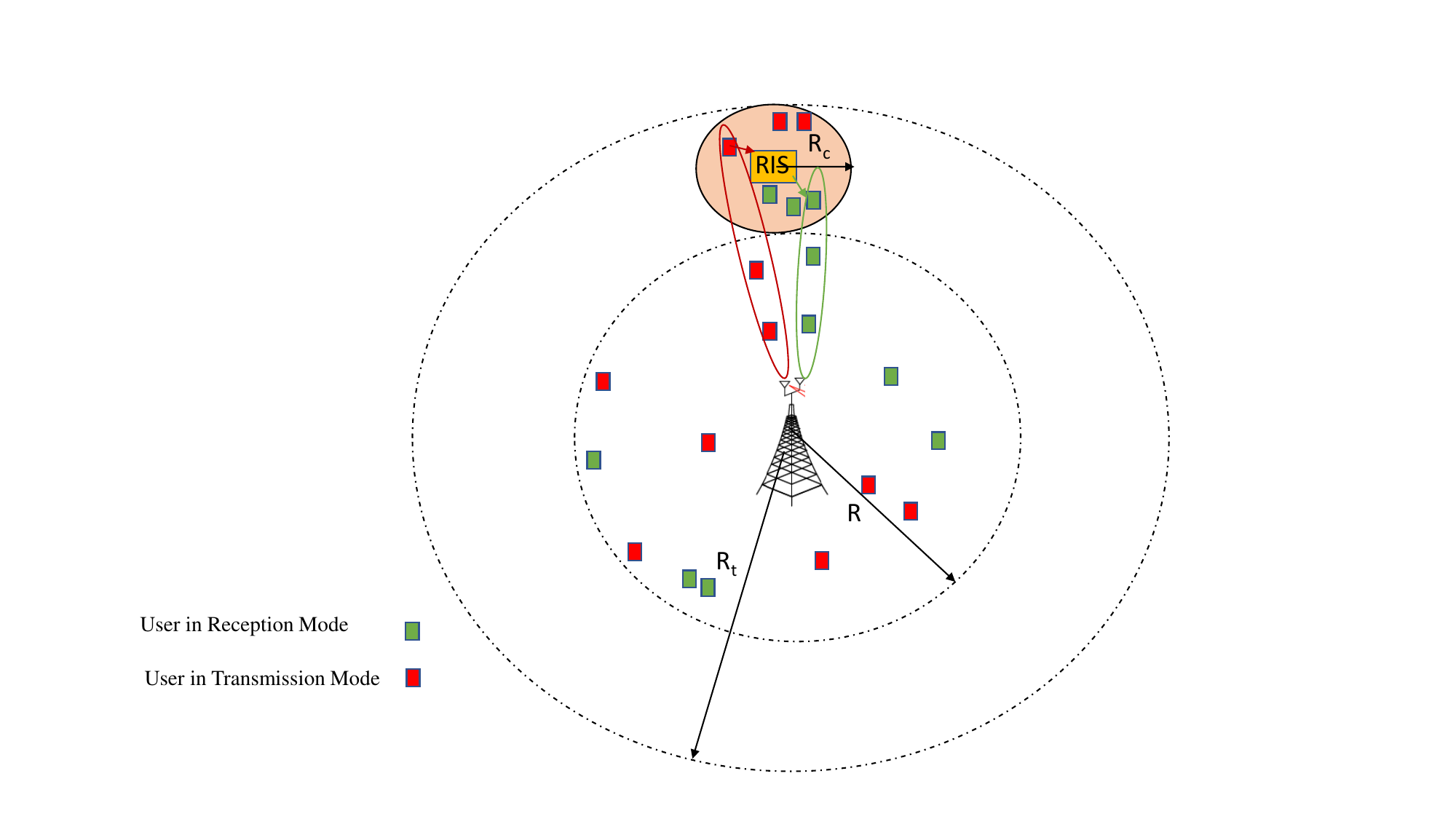}
\par\end{centering}

\protect\caption{\label{fig:System-Model}A STAR-RIS assisted FD communication system.}
\end{figure}

Considering a STAR-RIS-assisted multiple-users FD NOMA communication
system shown in Fig. \ref{fig:System-Model}. The FD-BS is situated
at the cell-center with coverage radius, $R_{t}$, and the STAR-RIS
with $N$ reconfigurable elements is situated at the cell-edge with
coverage radius $R_{r}$. According to this deployment, the total
coverage area $R_{t}$ can be divided into two areas, the cell-center
area with radios $R$, and the cell-edge area with radios $R_{r}$.
The FD-BS is equipped with two antennas, one for transmission and
one for reception, while the users are equipped with a single antenna.
Number of the DL and UL cell-center users are $K_{cd}\textrm{ and }K_{cu}$,
respectively, where $K_{cd}+K_{cu}=K_{c}$, and number of the DL and
UL cell-edge users are $K_{ed}\textrm{ and }K_{eu}$, respectively,
where $K_{ed}+K_{eu}=K_{e}$,  the total number of the DL and UL
users are $K_{cd}+K_{ed}=K_{d}$ and $K_{cu}+K_{eu}=K_{u}$, respectively.
The users are assumed to be uniformly distributed in the areas as
shown in Fig \ref{fig:System-Model}. Rician fading model is assumed
for the STAR-RIS related channels, while Rayleigh fading model is
assumed for BS to users and user to user channels due to extensive
scatterers. It is also assumed that the BS and the RIS can only know
the statistical CSI and the users' locations. 

In this work, the ES operating protocol is implemented at the STAR-RIS.
Thus, the STAR-RIS elements operate in the transmission ($t$) and
reflection ($r$) modes, and the energy of the incident signal on
each element is split into two portions, one for the transmitted signals
and one for the reflected signals with the energy-splitting coefficients
(amplitude coefficients) $\rho_{n}^{r}$ and $\rho_{n}^{t}$ ($\rho_{n}^{r}+\rho_{n}^{t}\leq1$).
The transmission and reflection matrices of the STAR-RIS can be written
as $\Theta_{k}=\textrm{diag}\left(\rho_{1}^{k}\theta_{1}^{k},....,\rho_{N}^{k}\theta_{N}^{k}\right),k\in\left\{ t,r\right\} $,
where $\theta_{n}^{k}=e^{j\phi_{n}^{k}}$, $\rho_{n}^{k}\in\left[0,1\right]$,
$\rho_{n}^{k}>1$ and $\left|\theta_{n}^{k}\right|=1$.

\subsection{Data Transmission and Problem Formulation}

In the considered model, the FD-BS employs NOMA scheme to serve the
DL and UL users. To reduce the detection complexity of NOMA technique,
the DL and UL users are divided into orthogonal clusters. Number of
the DL clusters, $1\leq M_{d}\leq\frac{K_{d}}{2}$, and UL clusters,
$1\leq M_{u}\leq\frac{K_{u}}{2}$. In addition, number of the users
in each DL cluster and UL cluster are denoted by $\kappa_{d}$ and
$\kappa_{u}$, respectively, where $2\leq\kappa_{d}\leq K_{d}$ and
$2\leq\kappa_{u}\leq K_{u}$. The joint user-clustering and STAR-RIS
design to maximize the sum rate can be formulated as

\begin{eqnarray}
 & \underset{\rho,\mathbf{\theta},\beta}{\max}\,\stackrel[j=1]{\frac{K_{d}}{2}}{\sum}\stackrel[i=1]{K_{d}}{\sum}\beta_{d_{i,j}}\bar{R}_{u_{i,jd}}+\stackrel[j=1]{\frac{K_{u}}{2}}{\sum}\stackrel[i=1]{K_{u}}{\sum}\beta_{u_{i,j}}\bar{R}_{u_{i,ju}}\nonumber \\
(C.1) & \textrm{s.t}\:\stackrel[j=1]{\frac{K_{d}}{2}}{\sum}\beta_{d_{i,j}}\bar{R}_{u_{i,jd}}>\hat{R}_{d_{i}},\:\stackrel[j=1]{\frac{K_{u}}{2}}{\sum}\beta_{u_{i,j}}\bar{R}_{u_{i,ju}}>\hat{R}_{u_{i}}\nonumber \\
(C.2) & R_{u_{kd}\rightarrow u_{id}}\geqslant R_{u_{id}},k>i\nonumber \\
(C.3) & \beta_{d_{i,j}}\in\left\{ 0,1\right\} ,\,\beta_{u_{i,j}}\in\left\{ 0,1\right\} ,\forall i,j\nonumber \\
(C.4) & \stackrel[j=1]{\frac{K_{d}}{2}}{\sum}\beta_{d_{i,j}}=1,\,\stackrel[j=1]{\frac{K_{u}}{2}}{\sum}\beta_{u_{i,j}}=1,\forall i\nonumber \\
(C.5) & 2\leq\stackrel[i=1]{K_{d}}{\sum}\beta_{d_{i,j}}\leq K_{d},\,2\leq\stackrel[i=1]{K_{u}}{\sum}\beta_{u_{i,j}}\leq K_{u},\forall j\nonumber \\
(C.6) & \left(\rho_{n}^{r}\right)+\left(\rho_{n}^{t}\right)=1,\forall n\in N\nonumber \\
(C.7) & \rho_{n}^{k}\geq0,\left|\theta_{n}^{k}\right|=1,\forall n\in N\label{eq:1}
\end{eqnarray}

\noindent where $\bar{R}_{u_{i,jd}}$ is the average rate of the $i^{th}$
DL user in cluster $j$, $\bar{R}_{u_{i,ju}}$ is the average rate
of the $i^{th}$ UL user in cluster $j$, and 

\noindent 
\[
\beta_{p_{i,j}}=\begin{cases}
1, & \textrm{if a user }i\textrm{ is grouped into cluster }j\\
0, & \textrm{otherwise}
\end{cases}
\]

\noindent The first constraint $\left(C.1\right)$ provides the data
rate requirements of the DL and UL users, and the second constraint
$(C.2)$ provides the SIC implementation condition, while the constraint
$\left(C.3\right)$ explains that $\beta_{d_{i,j}}$ and $\beta_{u_{i,j}}$
are integer variables. The constraints $\left(C.\text{4}\right)$
and $\left(C.\text{5}\right)$ are required to ensure that each user
is assigned to only one cluster and that each cluster has at least
two users, constraints $\left(C.6\right)$ and $\left(C.7\right)$
for the amplitude and phase shift on each STAR-RIS element.

As we can observe from (\ref{eq:1}), it is extremely difficult to
find the solution of the problem due to its non-convexity in nature.
Specifically, an exhaustive search algorithm is required to find the
optimal solution for NOMA user clustering scheme. Thus, for each DL
and UL user all possible combinations should be considered. For instance,
in the DL number of the possible combinations is: $\stackrel[i=1]{K_{d}}{\sum}\left(\begin{array}{c}
K_{d}\\
i
\end{array}\right)$. It is clear that, the computational complexity of the optimal user
clustering is extremely high and impractical with a large number of
users. In addition, considering the two variables, $\rho,\mathbf{\theta}$,
to obtain the optimal STAR-RIS design makes the problem more complicated.
In the following sections we divide the main problem into two sub-problems.
We first develop a less complex and efficient solution for user clustering
scheme, and then we consider the optimal system design.

\section{User Clustering Scheme \label{sec:Cell-edge-users-added}}

A user clustering scheme is an efficient solution to reduce the decoding
and implementation complexities of SIC in NOMA systems. In the user
clustering scheme number of the signals decoded by SIC can be reduced
by distributing the users into small groups \cite{maga,nearfar1,nearfar2}. 

In this section, we introduce low-complexity sub-optimal user clustering
schemes for a STAR-RIS-aided FD-NOMA systems. The proposed schemes
exploit the users locations, and the distances between the users and
the FD-BS.

\subsection{Key Principles}

In DL, following the principles of NOMA, the users are sorted in descending
order based on their distances to the BS, $l_{b,u_{Kd}}>..l_{b,u_{2d}}>l_{b,u_{1d}}$
where $l_{b,u_{kd}}$is the distance between the BS and user $k$.
After applying SIC, the rate of the closet users to the BS (the strong/cell-center
users) will depend essentially on their channel variances and the
amount of power allocated to each user. Thus, it is reasonable and
beneficial to distribute the strong users into different clusters
to enhance the achievable sum-rate. For the far users (the weak/cell-edge
users), it is more reasonable and useful to include them with the
strong users. This is because the strong users can achieve high rates
with low power portions, and thus large portion of power can be allocated
to the weak users. Accordingly, the key idea here is to group the
strongest users and the weakest users into the same cluster \cite{clust2,clus3,clusMAIN}. 

In UL, similarly, the users are sorted in descending order according
to their distances to the BS, $l_{b,u_{Ku}}>....>l_{b,u_{2u}}>l_{b,u_{1u}}$.
In UL NOMA, the strong users have less impact on the performance of
the weak users, because the strongest users' messages will be canceled/reduced
using SIC scheme. Thus, the users can transmit their messages using
the highest available power. Accordingly, it is useful to cluster
the strong users together to enhance the achievable sum-rate of the
cluster \cite{clust2,clus3,clusMAIN}. It is also worth mentioning
that, in UL the BS is the receiver, thus it can receive the messages
in any users order.

\subsection{User Clustering Algorithm}

To simplify NOMA operation, the cell-center users are virtually partitioned
into two groups, group 1 ($\mathcal{G}_{1}$ ) and group 2 ($\mathcal{G}_{2}$
), their locations are as much different as possible \cite{clust1,clust2,clust3}.
The users in $\mathcal{G}_{1}$ are the nearest to the BS and the
users in $\mathcal{G}_{2}$ are the farthest users to the BS. Number
of the users in group $\mathcal{G}_{1}$ is $K_{d_{1}}$ in DL and
$K_{u_{1}}$in UL, while number of the users in group $\mathcal{G}_{2}$
is $K_{d_{2}}$ in DL and $K_{u_{2}}$ in UL %
\footnote{Note that the analysis in this work is also applicable when the DL
and UL groups have different numbers of users%
}. This partitioning of the users results in three different groups
in total, $\mathcal{G}_{1}$, $\mathcal{G}_{2}$ and the cell-edge
users group $\mathcal{G}_{3}$. Based on the discussions above, we
sort the DL and UL users in each group in descending order, $l_{b,u_{K_{di}}}>...>l_{b,u_{2di}}>l_{b,u_{1di}}$
and $l_{b,u_{K_{ui}}}>...>l_{b,u_{2ui}}>l_{b,u_{1ui}}$ where $i\in\left\{ 1,2,3\right\} $
is the group index.

In DL, following the key principles, the strongest users in $\mathcal{G}_{1}$
and $\mathcal{G}_{2}$ and the weakest user in $\mathcal{G}_{3}$
are grouped into the same cluster, while the second strongest users
in $\mathcal{G}_{1}$ and $\mathcal{G}_{2}$ and the second weakest
user in $\mathcal{G}_{3}$ are grouped into another cluster, and so
on. 

In UL, as we explained above, the strongest users in each group are
included in a cluster, the second strongest users in each group are
grouped in another cluster, and so on.

\begin{figure}
\noindent \begin{centering}
\includegraphics[bb=190bp 100bp 750bp 500bp,clip,scale=0.6]{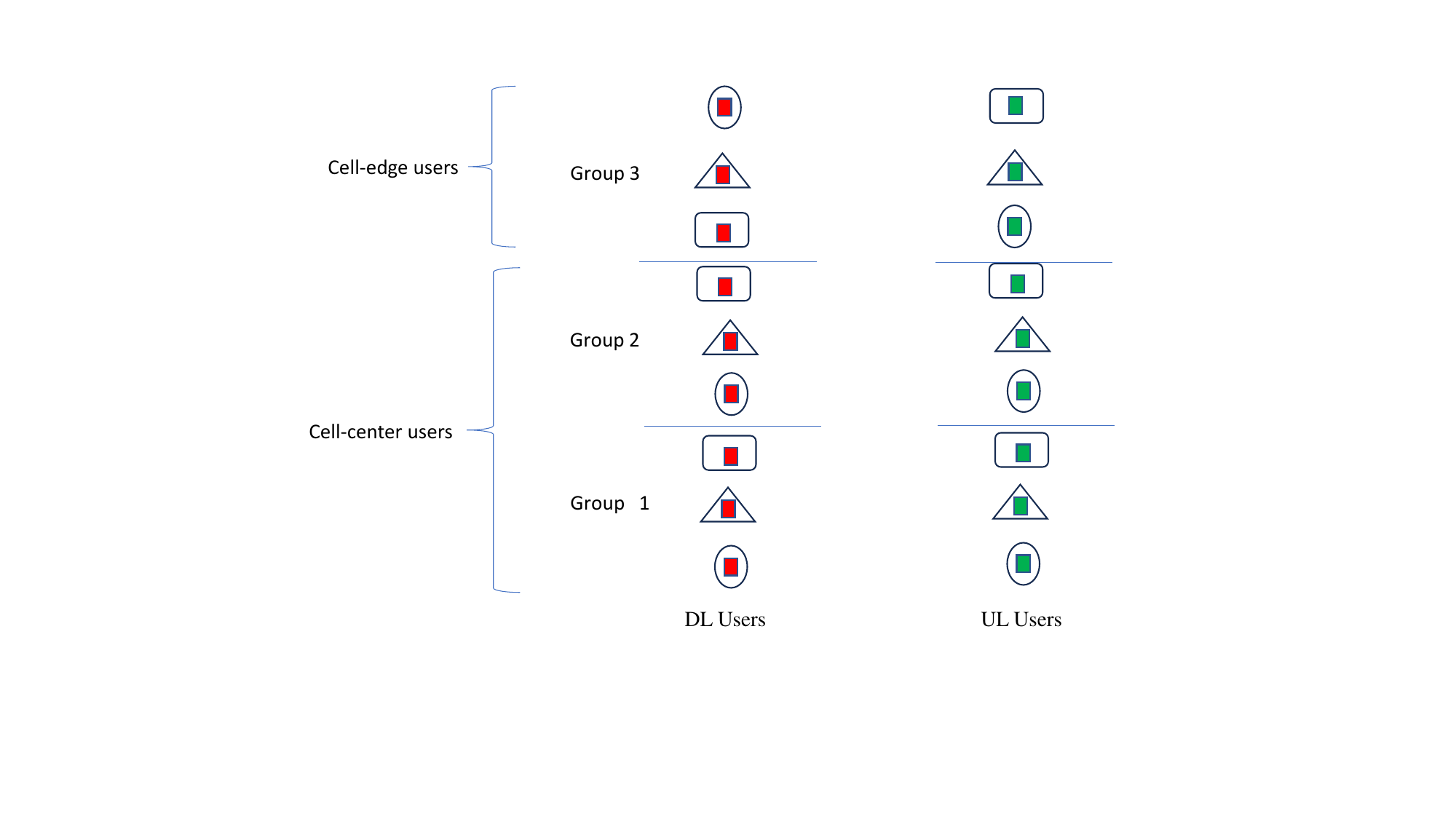}
\par\end{centering}

\protect\caption{\label{fig:Illustration}Illustration of user clustering scheme to
serve 6 cell-center users and 3 cell-edge users in DL and UL modes.}

\end{figure}

All steps of the proposed algorithm are summarized in Aalgorithm1.
In addition, for clarity in Fig. \ref{fig:Illustration}, we illustrate
the user clustering scheme for 6 cell-center users, and 3 cell-edge
users in DL and UL modes%
\footnote{Increasing number of the users in a cluster increases the detection
complexity but also reduces the usage of the resources/time slots.
Thus, moving a step above NOMA paring scheme will provide better usage
of the resources, which can compensate for the small errors in the
SIC stage.%
}.

\begin{algorithm}
1. Calculate the distances between the users and the BS.

2. Sort the DL and UL users according to their distances to the BS.

3. Divide the cell-center users into two groups, and assign the users
in each group: $\underset{\textrm{DL users in \ensuremath{\mathcal{G}_{1}}}}{\underbrace{l_{b,u_{1d1}}^{-m}>l_{b,u_{2d1}}^{-m}>.}.}.\underset{\textrm{DL users in \ensuremath{\mathcal{G}_{2}}}}{\underbrace{>l_{b,u_{1d2}}^{-m}>l_{b,u_{2d2}}^{-m}>.}}.$ 

$.\underset{\textrm{DL users in \ensuremath{\mathcal{G}_{3}}}}{.\underbrace{>l_{b,u_{K_{ed}-1}}^{-m}>l_{b,u_{K_{ed}}}^{-m}}}$and

$\underset{\textrm{UL users in \ensuremath{\mathcal{G}_{1}}}}{\underbrace{l_{b,u_{1u1}}^{-m}>l_{b,u_{2u1}}^{-m}>.}}.\underset{\textrm{UL users in \ensuremath{\mathcal{G}_{2}}}}{.\underbrace{>l_{b,u_{1u2}}^{-m}>l_{b,u_{2u2}}^{-m}>.}}$ 

$\underset{\textrm{UL users in \ensuremath{\mathcal{G}_{3}}}}{\underbrace{..>l_{b,u_{K_{eu}-1}}^{-m}>l_{b,u_{K_{eu}}}^{-m}}}$. 

4.1 Group the DL users into clusters: $\textrm{1st cluster}=\left\{ l_{b,u_{1d1}}^{-m},l_{b,u_{1d2}}^{-m},l_{b,u_{K_{ed}}}^{-m}\right\} $,
$\textrm{2nd cluster}=\left\{ l_{b,u_{2d1}}^{-m},l_{b,u_{2d2}}^{-m},l_{b,u_{K_{ed}-1}}^{-m}\right\} $...
and $\textrm{last cluster}=\left\{ l_{b,u_{K_{d1}}}^{-m},l_{b,u_{_{K_{d2}}}}^{-m},l_{b,u_{K_{d1}+K_{d2}+1}}^{-m}\right\} $. 

4.2 Group the UL users into clusters:$\textrm{1st cluster}=\left\{ l_{b,u_{1u1}}^{-m},l_{b,u_{1u2}}^{-m},l_{b,u_{1u3}}^{-m}\right\} $,
$\textrm{2nd cluster}=\left\{ l_{b,u_{2u1}}^{-m},l_{b,u_{2u2}}^{-m},l_{b,u_{2u3}}^{-m}\right\} $
and $\textrm{last cluster}=\left\{ l_{b,u_{K_{u1}}}^{-m},l_{b,u_{K_{u2}}}^{-m},l_{b,u_{K_{eu}}}^{-m}\right\} $ 

\protect\caption{User Clustering Algorithm.}

\end{algorithm}

\section{Ergodic Rate Analysis \label{sec:Ergodic-Rate-Analysis}}

In this section, we derive the ergodic rates of the DL and UL users,
which will be used later to find the optimal system design.

\subsection{DL}

In the DL transmission, the BS transmits the following signal to a
cluster in the system

\begin{equation}
s=\stackrel[i=1]{3}{\sum}\sqrt{\alpha_{i}}x_{u_{id}}
\end{equation}

\noindent where $\alpha_{i}$ is the power allocation coefficients
with $\alpha_{1}+\alpha_{2}+\alpha_{3}\leq1$ ,$\alpha_{1}<\alpha_{2}<\alpha_{3}$
and $x_{u_{id}}$ is the information signal of user $i$ with unit
variance.

\subsubsection{Cell-center, Group $\mathcal{G}_{1}$}

The received signal at the DL strongest user in a cluster (user 1)
can be written as

\begin{equation}
y_{u_{1d}}=\sqrt{P_{b}l_{b,u_{1d}}^{-m}}h_{b,u_{1d}}s+I_{u_{1d}}+n_{u_{1d}}
\end{equation}

\noindent where $P_{b}$ is the BS transmit power, $l_{b,u_{1d}}^{-m}$
is the path-loss between the BS and the user, $m$ is the path-loss
exponent, $h_{b,u_{1d}}\thicksim CN(0,1)$ is the channel between
the BS and the user, $n_{u_{1d}}$ is the additive white Gaussian
noise (AWGN) at the user, $n_{u_{1d}}\sim CN\left(0,\sigma_{u_{1d}}^{2}\right)$,
and $I_{u_{1d}}$ is the interference caused by the UL transmissions
and given by 

\[
I_{u_{1d}}=\underset{\textrm{UL cell-center user 1 signal}}{\underbrace{\sqrt{p_{u_{1u}}l_{u_{1d},u_{1u}}^{-m}}h_{u_{1d},u_{1u}}x_{u_{1u}}}}
\]

\[
+\underset{\textrm{ UL cell-center user 2}\textrm{ signal}}{\underbrace{\sqrt{p_{u_{2u}}l_{u_{1d},u_{2u}}^{-m}}h_{u_{1d},u_{2u}}x_{u_{2u}}}}
\]

\[
+\underset{\textrm{UL cell-edge user signal }}{\underbrace{\sqrt{p_{u_{3u}}l_{r,u_{3u}}^{-m}l_{r,u_{1d}}^{-m}}\mathbf{g}_{r,u_{1d}}\Theta_{t}\mathbf{g}_{r,u_{3u}}x_{u_{3u}}}}
\]

\noindent where $x_{u_{iu}}$ and $p_{u_{iu}}$ are the transmit power
and the signal of user $i$, $h_{u_{1d},u_{iu}}\thicksim CN(0,1)$
is the channel between the user and UL user $i$, $l_{u_{1d},u_{iu}}^{-m},l_{r,u_{3u}}^{-m},l_{r,u_{1d}}^{-m}$
represent the path-losses between, the user and UL user $i$, the
STAR-RIS and the UL cell-edge user, the STAR-RIS and the user, respectively,
and $\mathbf{g}_{_{r,u_{1d}}}=\left(\sqrt{\frac{\kappa_{r,u_{1d}}}{\kappa_{r,u_{1d}}+1}}\bar{\mathbf{g}}_{_{r,u_{1d}}}+\sqrt{\frac{1}{\kappa_{r,u_{1d}}+1}}\tilde{\mathbf{g}}_{_{r,u_{1d}}}\right),\mathbf{g}_{_{r,u_{3u}}}=\left(\sqrt{\frac{\kappa_{r,u_{3u}}}{\kappa_{r,u_{3u}}+1}}\bar{\mathbf{g}}_{_{r,u_{3u}}}+\sqrt{\frac{1}{\kappa_{r,u_{3u}}+1}}\tilde{\mathbf{g}}_{_{r,u_{3u}}}\right)$
are the channel vectors between the RIS and the user, and the RIS
and the UL cell-edge user, respectively. 

The SINR at the DL strongest user in a cluster, user 1, can be written
as 

\begin{equation}
\gamma_{u_{1d}}=\frac{\alpha_{1}P_{b}A_{u_{1d}}}{\Xi P_{b}\left(\alpha_{2}+\alpha_{3}\right)A_{u_{1d}}+B_{u_{1d}}+C_{u_{1d}}+D_{u_{1d}}+\sigma_{u_{1d}}^{2}}
\end{equation}

\noindent where

\noindent $A_{u_{1d}}=\left|\sqrt{l_{b,u_{1d}}^{-m}}h_{b,u_{1d}}\right|^{2}$,

\noindent $B_{u_{1d}}=p_{u_{1u}}\left|\sqrt{l_{u_{1d},u_{1u}}^{-m}}h_{u_{1d},u_{1u}}\right|^{2}$,

\noindent $C_{u_{1d}}=p_{u_{2u}}\left|\sqrt{l_{u_{1d},u_{2u}}^{-m}}h_{u_{1d},u_{2u}}\right|^{2}$,

\noindent $D_{u_{1d}}=p_{u_{3u}}l_{r,u_{3u}}^{-m}l_{r,u_{1d}}^{-m}\left|\mathbf{g}_{r,u_{1d}}\Theta_{t}\mathbf{g}_{r,u_{3u}}\right|^{2}$, 

\noindent and $0\leq\Xi\leq1$ is the SIC error factor which represents
a fraction of the power that remains as interference due to imperfect
SIC.
\begin{thm}
The ergodic rate of the DL strongest user in a cluster can be evaluated
by

\[
\mathcal{E}\left[R_{u_{1d}}\right]\approx\frac{1}{M_{d}}\log_{2}\left(1+\right.
\]

\begin{equation}
\left.\frac{P_{b_{1}}x_{1_{u_{1d}}}}{\Xi P_{b}\left(\alpha_{2}+\alpha_{3}\right)x_{1_{u_{1d}}}+p_{u_{1,2}}y_{1_{u_{1d}}}+p_{u_{3u}}\omega_{r,u_{3u}}^{r,u_{1d}}y_{2_{u_{1d}}}+\sigma_{u_{1d}}^{2}}\right)
\end{equation}

\noindent where $x_{1_{u_{1d}}},y_{1_{u_{1d}}},\textrm{ and }y_{2_{u_{1d}}}$
are defined as
\end{thm}
\[
x_{1_{u_{1d}}}=2a_{1_{u1d}}\Gamma\left(k_{cd_{1}}\right)\mathcal{H}\left(\left\{ k_{cd_{1}},\frac{1+m}{2},\frac{m}{2}\right\} ,\left\{ \frac{1}{2},1+K_{cd}\right\} ,R^{2}\right)
\]

\begin{equation}
-mRa_{1_{u1d}}\Gamma\left(\frac{1}{2}+k_{cd_{1}}\right)\mathcal{H}\left(\left\{ \frac{1}{2}+k_{cd_{1}},\frac{1+m}{2},\frac{2+m}{2}\right\} ,\left\{ \frac{3}{2},\frac{3}{2}+K_{cd}\right\} ,R^{2}\right)\label{eq:5}
\end{equation}

\[
y_{1_{u_{1d}}}=\frac{2}{\left(2-3m+m^{2}\right)R^{2}}-\frac{2\mathcal{H}\left(\left\{ \frac{1}{2},-1+\frac{m}{2},-\frac{1}{2}+\frac{m}{2}\right\} ,\left\{ \frac{-1}{2},1\right\} ,4R^{2}\right)}{\left(2-3m+m^{2}\right)R^{2}}
\]

\[
-\mathcal{H}\left(\left\{ \frac{3}{2},\frac{1}{2}+\frac{m}{2},\frac{m}{2}\right\} ,\left\{ \frac{1}{2},3\right\} ,4R^{2}\right)+\frac{64mR\,\mathcal{H}\left(\left\{ 2,\frac{1}{2}+\frac{m}{2},1+\frac{m}{2}\right\} ,\left\{ \frac{3}{2},\frac{7}{2}\right\} ,4R^{2}\right)}{15\pi}
\]

\begin{equation}
-\frac{64mR\,\mathcal{H}\left(\left\{ 2,\frac{1}{2}+\frac{m}{2},1+\frac{m}{2}\right\} ,\left\{ \frac{5}{2},\frac{5}{2}\right\} ,4R^{2}\right)}{9\pi}\label{eq:7}
\end{equation}

\[
y_{2_{u_{1d}}}=\left(2a_{1_{u3u}}\Gamma\left(k_{eu_{3}}\right)\mathcal{H}\left(\left\{ k_{eu_{3}},\frac{1+m}{2},\frac{m}{2}\right\} ,\left\{ \frac{1}{2},1+K_{eu}\right\} ,R_{r}^{2}\right)\right.
\]

\[
\left.-mR_{r}a_{1_{u3u}}\Gamma\left(\frac{1}{2}+k_{eu_{3}}\right)\mathcal{H}\left(\left\{ \frac{1}{2}+k_{eu_{3}},\frac{1+m}{2},\frac{2+m}{2}\right\} ,\left\{ \frac{3}{2},\frac{3}{2}+K_{eu}\right\} ,R_{r}^{2}\right)\right)
\]

\begin{equation}
\times\stackrel[j=1]{C}{\sum}\textrm{H}_{j}\,\left(1+\left(R\, r_{j}+R\right)\right)^{-m}\frac{2\left(R\, r_{j}+R\right)}{\pi R^{2}}\cos^{-1}\left(\frac{1}{\left(R\, r_{j}+R\right)}\left(r_{1}+\frac{\left(\left(R\, r_{j}+R\right)^{2}-r_{1}^{2}\right)}{2\left(R+r_{1}\right)}\right)\right)\label{eq:9}
\end{equation}

\noindent \emph{while $k_{cd_{1}}$ is the user order $1\leq k_{cd_{1}}\leq K_{d_{1}}$,
$\mathcal{H}\left(.\right)$ is the hypergeometric function, $p_{u_{1,2}}=\left(p_{u_{1u}}+p_{u_{2u}}\right)$,
$a_{1_{u1d}}=\frac{\sqrt{\pi}\Gamma\left(1-k_{cd_{1}}+K_{cd}\right)2K_{cd}!}{4\left(k_{cd_{1}}-1\right)!\left(K_{cd}-k_{cd_{1}}\right)!}$,
$a_{1_{u3u}=}\frac{\sqrt{\pi}\Gamma\left(1-k_{eu_{3}}+K_{eu}\right)2K_{eu}!}{4\left(k_{eu_{3}}-1\right)!\left(K_{eu}-k_{eu_{3}}\right)!}$,
$\omega_{r,u_{3u}}^{r,u_{1d}}=\left(\varpi_{r,u_{3u}}^{r,u_{1d}}\xi_{1}+\hat{\varpi}_{r,u_{3u}}^{r,u_{1d}}\right)$,
$\xi_{1}=\left|\bar{\mathbf{g}}_{_{r,u1d}}\Theta_{t}\mathbf{\bar{g}}_{r,u_{3u}}\right|^{2},\varpi_{y}^{x}=\frac{\kappa_{x}}{\kappa_{x}+1}\frac{\kappa_{y}}{\kappa_{y}+1}$,
$\hat{\varpi}_{y}^{x}=\frac{\kappa_{x}}{\kappa_{x}+1}\frac{\stackrel[n=1]{N}{\sum}\left|\rho_{n}^{k}\right|^{2}}{\kappa_{y}+1}+\frac{\kappa_{y}}{\kappa_{y}+1}\frac{\stackrel[n=1]{N}{\sum}\left|\rho_{n}^{k}\right|^{2}}{\kappa_{x}+1}+\frac{1}{\kappa_{x}+1}\frac{\stackrel[n=1]{N}{\sum}\left|\rho_{n}^{k}\right|^{2}}{\kappa_{y}+1}$.}
\begin{IEEEproof}
The proof is provided in Appendix A.
\end{IEEEproof}
.

\subsubsection{Cell-center, Group $\mathcal{G}_{2}$}

The received signal at the DL second strongest user in a cluster (user
2) can be expressed as

\begin{equation}
y_{u_{2d}}=\sqrt{P_{b}l_{b,u_{2d}}^{-m}}h_{b,u_{2d}}s+I_{u_{2d}}+n_{u_{2d}}
\end{equation}

\noindent where $I_{u_{2d}}$ is the interference caused by the UL
transmissions and given by 

\[
I_{u_{2d}}=\underset{\textrm{UL cell-center user 1 signal}}{\underbrace{\sqrt{p_{u_{1u}}l_{u_{2d},u_{1u}}^{-m}}h_{u_{2d},u_{1u}}x_{u_{1u}}}}
\]

\[
+\underset{\textrm{UL cell-center user 2 signal}}{\underbrace{\sqrt{p_{u_{2u}}l_{u_{2d},u_{2u}}^{-m}}h_{u_{2d},u_{2u}}x_{u_{2u}}}}
\]

\[
+\underset{\textrm{ UL cell-edge user signal }}{\underbrace{\sqrt{p_{u_{3u}}l_{r,u_{3u}}^{-m}l_{r,u_{2d}}^{-m}}\mathbf{g}_{r,u_{2d}}\Theta_{t}\mathbf{g}_{r,u_{3u}}x_{u_{3u}}}}
\]

Thus, the SINR at the user can be written as 

\begin{equation}
\gamma_{u_{2d}}=\frac{\alpha_{2}P_{b}A_{u_{2d}}}{P_{b}A_{u_{2d}}\left(\Xi\alpha_{3}+\alpha_{1}\right)+B_{u_{2d}}+C_{u_{2d}}+D_{u_{2d}}+\sigma_{u_{2d}}^{2}}
\end{equation}

\noindent where

\noindent $A_{u_{2d}}=\left|\sqrt{l_{b,u_{2d}}^{-m}}h_{b,u_{2d}}\right|^{2}$,

\noindent $B_{u_{2d}}=p_{u_{1u}}\left|\sqrt{l_{u_{2d},u_{1u}}^{-m}}h_{u_{2d},u_{1u}}\right|^{2}$

\noindent $C_{u_{2d}}=p_{u_{2u}}\left|\sqrt{l_{u_{2d},u_{2u}}^{-m}}h_{u_{2d},u_{2u}}\right|^{2}$,

\noindent $D_{u_{2d}}=p_{u_{3u}}l_{r,u3u}^{-m}l_{r,u2d}^{-m}\left|\mathbf{g}_{r,u_{2d}}\Theta_{t}\mathbf{g}_{r,u_{3u}}\right|^{2}$.
\begin{thm}
The ergodic rate of the DL second strongest user in a cluster can
be evaluated by

\[
\mathcal{E}\left[R_{u_{2d}}\right]\approx\frac{1}{M_{d}}\log_{2}\left(1+\right.
\]

\begin{equation}
\left.\frac{P_{b_{2}}x_{1_{u_{2d}}}}{P_{b}x_{1_{u_{2d}}}\left(\Xi\alpha_{3}+\alpha_{1}\right)+p_{u_{1,2}}y_{1_{u_{2d}}}+p_{u_{3u}}\omega_{r,u_{3u}}^{r,u_{2d}}y_{2_{u_{2d}}}+\sigma_{u_{2d}}^{2}}\right)
\end{equation}

\noindent where $x_{1_{u_{2d}}}$ is defined as
\end{thm}
\[
x_{1_{u_{2d}}}=2a_{1_{u2d}}\Gamma\left(k_{cd_{2}}\right)\mathcal{H}\left(\left\{ k_{cd_{2}},\frac{1+m}{2},\frac{m}{2}\right\} ,\left\{ \frac{1}{2},1+K_{cd}\right\} ,R^{2}\right)
\]

\begin{equation}
-mRa_{1_{u2d}}\Gamma\left(\frac{1}{2}+k_{cd_{2}}\right)\mathcal{H}\left(\left\{ \frac{1}{2}+k_{cd_{2}},\frac{1+m}{2},\frac{2+m}{2}\right\} ,\left\{ \frac{3}{2},\frac{3}{2}+K_{cd}\right\} ,R^{2}\right)\label{eq:43}
\end{equation}

\noindent \emph{while $y_{1_{u_{2d}}}=y_{1_{u_{1d}}}$, $y_{2_{u_{2d}}}=y_{2_{u_{1d}}}$,
$k_{cd_{2}}$ is the user order $K_{d_{1}}+1\leq k_{cd_{2}}\leq K_{cd}$,
$\omega_{r,u_{3u}}^{r,u_{2d}}=\left(\varpi_{r,u_{3u}}^{r,u_{2d}}\xi_{2}+\hat{\varpi}_{r,u_{3u}}^{r,u_{2d}}\right)$,
$\xi_{2}=\left|\bar{\mathbf{g}}_{_{r,u2d}}\Theta_{t}\mathbf{\bar{g}}_{r,u_{3u}}\right|^{2}$
and $a_{1_{u2d}}=\frac{\sqrt{\pi}\Gamma\left(1-k_{cd_{2}}+K_{cd}\right)2K_{cd}!}{4\left(k_{cd_{2}}-1\right)!\left(K_{cd}-k\right)!}$.}
\begin{IEEEproof}
The proof is provided in Appendix B.
\end{IEEEproof}

\subsubsection{Cell-edge, Group $\mathcal{G}_{3}$}

The received signal at the cell-edge user, user 3, can be expressed
as

\begin{equation}
y_{u_{3d}}=\sqrt{P_{b}l_{b,r}^{-m}l_{r,u_{3d}}^{-m}}g_{r,u_{3d}}\Theta_{r}g_{b,r}s+I_{u_{3d}}+n_{u_{3d}}
\end{equation}

\noindent where $I_{u_{3d}}$ is the interference caused by the UL
transmissions and given by 

\[
I_{u_{3d}}=\underset{\textrm{UL cell-center user 1 signal}}{\underbrace{\sqrt{p_{u_{1u}}l_{r,u_{3d}}^{-m}l_{r,u_{1u}}^{-m}}\mathbf{g}_{r,u_{3d}}\Theta_{r}\mathbf{g}_{r,u_{1u}}x_{u_{1u}}}}
\]

\[
+\underset{\textrm{UL cell-center user 2 signal}}{\underbrace{\sqrt{p_{u_{2u}}l_{r,u_{3d}}^{-m}l_{r,u_{2u}}^{-m}}\mathbf{g}_{r,u_{3d}}\Theta_{r}\mathbf{g}_{r,u_{2u}}x_{u_{2u}}}}
\]

\[
+\underset{\textrm{UL cell-edge user signal }}{\underbrace{\sqrt{p_{u_{3u}}l_{r,u_{3d}}^{-m}l_{r,u_{3u}}^{-m}}\mathbf{g}_{r,u_{3d}}\Theta_{r}\mathbf{g}_{r,u_{3u}}x_{u_{3u}}}}
\]

The SINR at the user can be written as

\begin{equation}
\gamma_{u_{3d}}=\frac{A_{u_{3d}}}{P_{b_{1,2}}A_{u_{3d}}+B_{u_{3d}}+C_{u_{3d}}+D_{u_{3d}}+\sigma_{u_{3d}}^{2}}
\end{equation}

\noindent where $A_{u_{3d}}=\alpha_{3}P_{b}l_{b,r}^{-m}l_{r,u_{3d}}^{-m}\left|\mathbf{g}_{r,u_{3d}}\Theta_{r}\mathbf{g}_{b,r}\right|^{2}$,
$P_{b_{1,2}}=\left(\alpha_{1}+\alpha_{2}\right)P_{b}$,

\noindent $B_{u_{3d}}=p_{u_{1u}}l_{r,u_{3d}}^{-m}l_{r,u_{1u}}^{-m}\left|\mathbf{g}_{r,u_{3d}}\Theta_{r}\mathbf{g}_{r,u_{1u}}\right|^{2}$,

\noindent $C_{u_{3d}}=p_{u_{2u}}l_{r,u_{3d}}^{-m}l_{r,u_{2u}}^{-m}\left|\mathbf{g}_{r,u_{3d}}\Theta_{r}\mathbf{g}_{r,u_{2u}}\right|^{2}$,

\noindent $D_{u_{3d}}=p_{u_{3u}}l_{r,u_{3d}}^{-m}l_{r,u_{3u}}^{-m}\left|\mathbf{g}_{r,u_{3d}}\Theta_{r}\mathbf{g}_{r,u_{3u}}\right|^{2}$.
\begin{thm}
The ergodic rate of the DL weakest user in a cluster can be evaluated
by

\[
\mathcal{E}\left[R_{u_{3d}}\right]\approx\frac{1}{M_{d}}\log_{2}\left(1+\right.
\]

\begin{equation}
\left.\frac{P_{b_{3}}x_{u_{3d}}}{P_{b_{1,2}}x_{u_{3d}}+b_{1}y_{1_{u_{3d}}}+b_{2}y_{2_{u_{3d}}}+\sigma_{u_{3d}}^{2}}\right)
\end{equation}

\noindent where $x_{u_{3d}}=l_{b,r}^{-m}\omega_{b,r}^{r,u_{3d}}x_{1_{u_{3d}}}$,~$b_{1}=p_{u_{1u}}\omega_{r,u_{1u}}^{r,u_{3d}}+p_{u_{2u}}\omega_{r,u_{2u}}^{r,u_{3d}}$,
$b_{2}=p_{u_{3u}}\omega_{r,u_{3u}}^{r,u_{3d}}$, $x_{1_{u_{3d}}},y_{1_{u_{3d}}},y_{2_{u_{3d}}}$
are defined as
\end{thm}
\[
x_{1_{u_{3d}}}=2a_{1_{u3d}}\Gamma\left(k{}_{ed_{3}}\right)\mathcal{H}\left(\left\{ k{}_{ed_{3}},\frac{1+m}{2},\frac{m}{2}\right\} ,\left\{ \frac{1}{2},1+K_{ed}\right\} ,R_{r}^{2}\right)
\]

\begin{equation}
-mR_{r}a_{1_{u3d}}\Gamma\left(\frac{1}{2}+k{}_{ed_{3}}\right)\mathcal{H}\left(\left\{ \frac{1}{2}+k{}_{ed_{3}},\frac{1+m}{2},\frac{2+m}{2}\right\} ,\left\{ \frac{3}{2},\frac{3}{2}+K_{ed}\right\} ,R_{r}^{2}\right)\label{eq:52}
\end{equation}

\begin{equation}
y_{1_{u_{3d}}}=x_{1_{u_{3d}}}\stackrel[j=1]{C}{\sum}\textrm{H}_{j}\,\left(1+\left(R\, r_{j}+R\right)\right)^{-m}\frac{2\left(R\, r_{j}+R\right)}{\pi R^{2}}\cos^{-1}\left(\frac{1}{\left(R\, r_{j}+R\right)}\left(r_{1}+\frac{\left(\left(R\, r_{j}+R\right)^{2}-r_{1}^{2}\right)}{2\left(R+r_{1}\right)}\right)\right)
\end{equation}

\[
y_{2_{u_{3d}}}=x_{1_{u_{3d}}}a_{2_{u3d}}2\Gamma\left(k{}_{eu_{1}}\right)\mathcal{H}\left(\left\{ k{}_{eu_{1}},\frac{1+m}{2},\frac{m}{2}\right\} ,\left\{ \frac{1}{2},1+K_{eu}\right\} ,R_{r}^{2}\right)
\]

\begin{equation}
-mR_{r}x_{1_{u_{3d}}}a_{2_{u3d}}\Gamma\left(\frac{1}{2}+k{}_{eu_{1}}\right)\mathcal{H}\left(\left\{ \frac{1}{2}+k{}_{eu_{1}},\frac{1+m}{2},\frac{2+m}{2}\right\} ,\left\{ \frac{3}{2},\frac{3}{2}+K_{eu}\right\} ,R_{r}^{2}\right)\label{eq:56}
\end{equation}

\noindent \emph{while $k{}_{ed_{3}}$ is the user order $1\leq k{}_{ed_{3}}\leq K_{ed}$,
$\omega_{b,r}^{r,u_{3d}}=\varpi_{b,r}^{r,u_{3d}}\xi_{3}+\hat{\varpi}_{b,r}^{r,u_{3d}}$
$\omega_{r,u_{1u}}^{r,u_{3d}}=\varpi_{r,u_{1u}}^{r,u_{3d}}\xi_{4}+\hat{\varpi}_{r,u_{1u}}^{r,u_{3d}}$,
$\omega_{r,u_{2u}}^{r,u_{3d}}=\varpi_{r,u_{2u}}^{r,u_{3d}}\xi_{5}+\hat{\varpi}_{r,u_{2u}}^{r,u_{3d}}$
$\omega_{r,u_{3u}}^{r,u_{3d}}=\varpi_{r,u_{3u}}^{r,u_{3d}}\xi_{6}+\hat{\varpi}_{r,u_{3u}}^{r,u_{3d}}$
$\xi_{3}=\left|\mathbf{\bar{g}}_{_{r,u3d}}\Theta_{r}\bar{\mathbf{g}}_{b,r}\right|^{2}$,
$\xi_{4}=\left|\mathbf{\bar{g}}_{_{r,u3d}}\Theta\bar{\mathbf{g}}_{r,u_{1u}}\right|^{2}$,
$\xi_{5}=\left|\mathbf{\bar{g}}_{_{r,u3d}}\Theta\bar{\mathbf{g}}_{r,u_{2u}}\right|^{2}$,
$\xi_{6}=\left|\mathbf{\bar{g}}_{_{r,u3d}}\Theta\bar{\mathbf{g}}_{r,u_{3u}}\right|^{2}$,
$a_{1_{u3d}}=\frac{\sqrt{\pi}\Gamma\left(1-k{}_{ed_{3}}+K_{ed}\right)2K_{ed}!}{4\left(k{}_{ed_{3}}-1\right)!\left(K_{ed}-k{}_{ed_{3}}\right)!}$,
and $a_{2_{u3d}}=\frac{\sqrt{\pi}\Gamma\left(1-k{}_{eu_{1}}+K_{eu}\right)2K_{eu}!}{4\left(k{}_{eu_{1}}-1\right)!\left(K_{eu}-k{}_{eu_{1}}\right)!}$.}
\begin{IEEEproof}
The proof is provided in Appendix C.
\end{IEEEproof}

\subsection{UL}

In UL mode, the received signal at the BS can be expressed as

\[
y_{b}=\sqrt{p_{u_{1u}}l_{b,u_{1u}}^{-m}}h_{b,u_{1u}}x_{u_{1u}}+\sqrt{p_{u_{2u}}l_{b,u_{2u}}^{-m}}h_{b,u_{2u}}x_{u_{2u}}+
\]

\begin{equation}
\sqrt{p_{u_{3u}}l_{b,r}^{-m}l_{r,u_{3u}}^{-m}}\mathbf{g}_{b,r}\Theta_{t}\mathbf{g}_{r,u_{3u}}x_{u_{3u}}+I_{b}+n_{b}
\end{equation}

\noindent where $n_{b}$ is the AWGN at the BS, $n_{b}\sim CN\left(0,\sigma_{b}^{2}\right)$
and $I_{b}$ is the interference term which can be represented by 

\[
I_{b}=\underset{\textrm{self interference }}{\underbrace{\sqrt{P_{b}}h_{b,b}s}}+\underset{\textrm{reflection of the DL signal}}{\underbrace{\sqrt{P_{b}l_{b,r}^{-m}l_{b,r}^{-m}}\mathbf{g}_{b,r}^{H}\Theta_{t}\mathbf{g}_{b,r}s}}
\]

Several self interference suppression (SIS) techniques have been proposed
in the literature, such as passive cancellation, analog and digital
cancellations, etc \cite{SI1,SI2}. Employing these interference suppression
schemes will minimize the self interference to the background noise
floor. Following the results in \cite{SI1,SI2}, the residual self-interference
is assumed to follow Gaussian distribution with zero-mean and variance
$V$, $\tilde{s}\sim CN\left(0,V\right)$. Thus, the interference
term $I_{b}$ can be written as 

\[
I_{b}=\underset{\textrm{residual self interference }}{\underbrace{\tilde{s}}}+\underset{\textrm{reflection of the DL signal}}{\underbrace{\sqrt{P_{b}l_{b,r}^{-m}l_{b,r}^{-m}}\mathbf{g}_{b,r}^{H}\Theta_{t}\mathbf{g}_{b,r}s}}
\]

\noindent According to the experimental results, the variance of the
residual self interference can be modeled as $V=\beta P_{b}^{\lambda}$,
where $\beta$ and $\lambda$ ($0\le\lambda\le1$) are constants reflect
the quality of the cancellation technique.

\subsubsection{Cell-center, Group $\mathcal{G}_{1}$}

The SINR to detect the signal of the UL strongest user in a cluster,
user 1, can be written as,

\begin{equation}
\gamma_{u_{1u}}=\frac{p_{u_{1u}}A_{u_{1u}}}{p_{u_{2u}}B_{u_{1u}}+p_{u_{3u}}C_{u_{1u}}+P_{b}D_{u_{1u}}+V+\sigma_{u_{1u}}^{2}}
\end{equation}

\noindent where

\noindent $A_{u_{1u}}=\left|\sqrt{l_{b,u_{1u}}^{-m}}h_{b,u_{1u}}\right|^{2}$,

\noindent $B_{u_{1u}}=\left|\sqrt{l_{b,u_{2u}}^{-m}}h_{b,u_{2u}}\right|^{2}$

\noindent $C_{u_{1u}}=l_{b,r}^{-m}l_{r,u_{3u}}^{-m}\left|\mathbf{g}_{b,r}\Theta_{t}\mathbf{g}_{r,u_{3u}}\right|^{2}$

\noindent $D_{u_{1u}}=l_{b,r}^{-m}l_{b,r}^{-m}\left|\mathbf{g}_{b,r}^{H}\Theta_{t}\mathbf{g}_{b,r}\right|^{2}$.
\begin{thm}
The ergodic rate of the UL strongest user in a cluster can be evaluated
by

\[
\mathcal{E}\left[R_{u_{1u}}\right]\approx\frac{1}{M_{u}}\log_{2}\left(1+\right.
\]

\begin{equation}
\left.\frac{p_{u_{1u}}\chi_{u_{1u}}}{p_{u_{2u}}\chi_{u_{2u}}+p_{u_{3u}}\omega_{b,r}^{r,u_{3u}}y_{2_{u_{1u}}}+P_{b}y_{3_{u_{1u}}}+V+\sigma_{b}^{2}}\right)
\end{equation}

\noindent where $\chi_{u_{iu}},y_{2_{u_{1u}}}$ and $y_{3_{u_{1u}}}$
are defined as
\end{thm}
\[
\chi_{u_{iu}}=2a_{i_{u1u}}\Gamma\left(k_{cu_{i}}\right)\mathcal{H}\left(\left\{ k_{cu_{i}},\frac{1+m}{2},\frac{m}{2}\right\} ,\left\{ \frac{1}{2},1+K_{cu}\right\} ,R^{2}\right)
\]

\begin{equation}
-mRa_{i_{u1u}}\Gamma\left(\frac{1}{2}+k_{cu_{i}}\right)\mathcal{H}\left(\left\{ \frac{1}{2}+k_{cu_{i}},\frac{1+m}{2},\frac{2+m}{2}\right\} ,\left\{ \frac{3}{2},\frac{3}{2}+K_{cu}\right\} ,R^{2}\right)\label{eq:57}
\end{equation}

\[
y_{2_{u_{1u}}}=2a_{3_{u1u}}\Gamma\left(k_{eu_{1}}\right)\mathcal{H}\left(\left\{ k_{eu_{1}},\frac{1+m}{2},\frac{m}{2}\right\} ,\left\{ \frac{1}{2},1+K_{eu}\right\} ,R_{r}^{2}\right)
\]

\begin{equation}
-mR_{r}a_{3_{u1u}}\Gamma\left(\frac{1}{2}+k_{eu_{1}}\right)\mathcal{H}\left(\left\{ \frac{1}{2}+k_{eu_{1}},\frac{1+m}{2},\frac{2+m}{2}\right\} ,\left\{ \frac{3}{2},\frac{3}{2}+K_{eu}\right\} ,R_{r}^{2}\right)
\end{equation}

\[
y_{3_{u_{1u}}}=l_{b,r}^{-m}l_{b,r}^{-m}\left(\left(\frac{\kappa_{b,r}}{\kappa_{b,r}+1}\right)^{2}\xi_{8}+2\frac{\kappa_{b,r}}{\kappa_{b,r}+1}\frac{1}{\kappa_{b,r}+1}\stackrel[n=1]{N}{\sum}\left|\rho_{n}^{k}\right|^{2}\right.+\left(\frac{1}{\kappa_{b,r}+1}\right)^{2}
\]

\begin{equation}
\left(2\stackrel[n=1]{N}{\sum}\left|\rho_{n}^{k}\right|^{2}+\stackrel[n_{1}=1]{N}{\sum}\stackrel[n_{2}\neq n_{1}]{N}{\sum}\left(\rho_{n_{1}}^{k}e^{j\phi_{n_{1}}^{k}}\right)\left(\rho_{n_{2}}^{k}e^{j\phi_{n_{2}}^{k}}\right)^{H}\right)\left.+2\frac{\kappa_{b,r}}{\kappa_{b,r}+1}\frac{1}{\kappa_{b,r}+1}\left(\hat{\zeta}\stackrel[n=1]{N}{\sum}\left(\rho_{n}^{k}e^{j\phi_{n}^{k}}\right)^{H}\right)\right)\label{eq:61}
\end{equation}

\noindent \emph{while $k_{cu_{1}}$is the strongest cell-center user
order $1\leq k_{cu_{1}}\leq K_{u_{1}}$, $k_{cu_{2}}$is the second
strongest cell-center user order $K_{u_{1}}+1\leq k_{cu_{2}}\leq K_{cu}$,
$k{}_{eu_{1}}$ is the UL cell-edge user order $1\leq k{}_{eu_{1}}\leq K_{eu}$,
$\omega_{b,r}^{r,u_{3u}}=\varpi_{b,r}^{r,u_{3u}}\xi_{7}+\hat{\varpi}_{b,r}^{r,u_{3u}}$,
$a_{1_{u1u}}=\frac{\sqrt{\pi}\Gamma\left(1-k_{cu_{1}}+K_{cu}\right)2K_{cu}!}{4\left(k_{cu_{1}}-1\right)!\left(K_{cu}-k\right)!}$,
$a_{2_{u1u}}=\frac{\sqrt{\pi}\Gamma\left(1-k_{cu_{2}}+K_{cu}\right)2K_{cu}!}{4\left(k_{cu_{2}}-1\right)!\left(K_{cu}-k_{cu_{2}}\right)!}$,
$a_{3_{u1u}}=\frac{\sqrt{\pi}\Gamma\left(1-k_{eu_{1}}+K_{eu}\right)2K_{eu}!}{4\left(k_{eu_{1}}-1\right)!\left(K_{eu}-k_{eu_{1}}\right)!}$,
$\xi_{7}=\left|\bar{\mathbf{g}}_{b,r}\Theta_{t}\mathbf{\bar{g}}_{_{_{r,u3u}}}\right|^{2}$,
$\xi_{8}=\left|\mathbf{\bar{g}}_{b,r}^{H}\Theta_{t}\bar{\mathbf{g}}_{b,r}\right|^{2}$
and $\hat{\zeta}=\mathbf{\bar{g}}_{b,r}^{H}\Theta_{t}\bar{\mathbf{g}}_{b,r}$.}
\begin{IEEEproof}
The proof is provided in Appendix D.
\end{IEEEproof}

\subsubsection{Cell-center, Group $\mathcal{G}_{2}$}

The SINR to detect the UL second strongest user signal can be written
as,

\begin{equation}
\gamma_{u_{2u}}=\frac{p_{u_{2u}}A_{u_{2u}}}{\Xi p_{u_{1u}}B_{u_{2u}}+p_{u_{3u}}C_{u_{2u}}+P_{b}D_{u_{2u}}+V+\sigma_{b}^{2}}
\end{equation}

\noindent where

\noindent $A_{u_{2u}}=\left|\sqrt{l_{b,u_{2u}}^{-m}}h_{b,u_{2u}}\right|^{2}$, 

\noindent $B_{u_{2u}}=\left|\sqrt{l_{b,u_{1u}}^{-m}}h_{b,u_{1u}}\right|^{2}$

\noindent $C_{u_{2u}}=l_{b,r}^{-m}l_{r,u_{3u}}^{-m}\left|\mathbf{g}_{b,r}\Theta_{t}\mathbf{g}_{r,u_{3u}}\right|^{2}$, 

\noindent $D_{u_{2u}}=l_{b,r}^{-m}l_{b,r}^{-m}\left|\mathbf{g}_{b,r}^{H}\Theta_{t}\mathbf{g}_{b,r}\right|^{2}$.
\begin{thm}
The ergodic rate of the UL second strongest user in a cluster can
be evaluated by

\[
\mathcal{E}\left[R_{u_{2u}}\right]\approx\frac{1}{M_{u}}\log_{2}\left(1+\right.
\]

\begin{equation}
\left.\frac{p_{u_{2u}}\chi_{u_{2u}}}{\Xi p_{u_{1u}}\chi_{u_{1u}}+p_{u_{3u}}\omega_{b,r}^{r,u_{3u}}y_{2_{u_{2u}}}+P_{b}y_{3_{u_{2u}}}+V+\sigma_{b}^{2}}\right)
\end{equation}

\noindent where $\chi_{u_{iu}}$, $y_{2_{u_{2u}}}=y_{2_{u1_{2u}}}$
and $y_{3_{u_{2u}}}=y_{3_{u_{1u}}}$ are defined in (\ref{eq:57})-(\ref{eq:61}).\end{thm}
\begin{IEEEproof}
The proof is provided in Appendix D.
\end{IEEEproof}

\subsubsection{Cell-edge, Group $\mathcal{G}_{3}$ }

The SINR to detect the UL cell-edge user signal can be written as,

\begin{equation}
\gamma_{u_{3u}}=\frac{p_{u_{3u}}A_{u_{3u}}}{\Xi p_{u_{1u}}B_{u_{3u}}+\Xi p_{u_{2u}}C_{u_{3u}}+P_{b}D_{u_{3u}}+V+\sigma_{b}^{2}}
\end{equation}

\noindent where $A_{u_{3u}}=l_{b,r}^{-m}l_{r,u_{3u}}^{-m}\left|\mathbf{g}_{b,r}\Theta_{t}\mathbf{g}_{r,u_{3u}}\right|^{2}$,

\noindent $B_{u_{3u}}=\left|\sqrt{l_{b,u_{1u}}^{-m}}h_{b,u_{1u}}\right|^{2}$

\noindent $C_{u_{3u}}=\left|\sqrt{l_{b,u_{2u}}^{-m}}h_{b,u_{2u}}\right|^{2}$

\noindent $D_{u_{3u}}=l_{b,r}^{-m}l_{b,r}^{-m}\left|\mathbf{g}_{b,r}^{H}\Theta\mathbf{g}_{b,r}\right|^{2}$.
\begin{thm}
The ergodic rate of the UL weakest user in a cluster can be evaluated
by

\[
\mathcal{E}\left[R_{u_{3u}}\right]\approx\frac{1}{M_{u}}\log_{2}\left(1+\right.
\]

\begin{equation}
\left.\frac{p_{u_{3u}}\omega_{b,r}^{r,u_{3u}}x_{1_{u_{3u}}}}{\Xi p_{u_{1u}}\chi_{u_{1u}}+\Xi p_{u_{2u}}\chi_{u_{2u}}+P_{b}y_{3_{u_{3u}}}+V+\sigma_{b}^{2}}\right)
\end{equation}

\noindent where $\chi_{u_{iu}}$, $x_{1_{u_{3u}}}=y_{2_{u_{1u}}}$
, and $y_{3_{u_{3u}}}=y_{3_{u_{1u}}}$ are defined in (\ref{eq:57})-(\ref{eq:61}).\end{thm}
\begin{IEEEproof}
The proof is provided in Appendix D.
\end{IEEEproof}

\section{Systems Design\label{sec:System-Design}}

As we can observe from the Theorems, the ergodic rates depend on the
amplitudes and phase shifts of the STAR-RIS elements and also on the
transmit powers. Thus, these parameters are optimized in this Section
to maximize the weighted sum rate of a DL cluster and an UL cluster.

\subsection{Simultaneous Amplitudes and Phase-Shifts Optimization}

The total weighted sum-rate at a given time slot and power transmission
can be optimized by considering the following problem 

\begin{eqnarray}
 & \underset{\rho,\mathbf{\theta}}{\max}\,\stackrel[i=1]{3}{\sum}\varpi_{u_{id}}\bar{R}_{u_{id}}+\stackrel[i=1]{3}{\sum}\varpi_{u_{iu}}\bar{R}_{u_{iu}}\nonumber \\
 & \textrm{s.t}\:\left(\rho_{n}^{r}\right)+\left(\rho_{n}^{t}\right)=1,\forall n\in N\nonumber \\
 & \rho_{n}^{k}\geq0,\left|\theta_{n}^{k}\right|=1,\forall n\in N\label{eq:29}
\end{eqnarray}

\noindent where $\varpi_{u_{id}}$ and $\varpi_{u_{iu}}$ are the
weighting factors, which signify the priority assigned to each user
and $\mathbf{\mathbf{\theta}}=\left[\theta^{t},\theta^{r}\right]$,
$\mathbf{\mathbf{\mathbf{\rho}}}=\left[\rho^{t},\rho^{r}\right]$.
The non-convexity of the problem in (\ref{eq:29}) with the two optimization
variables make the problem challenging and hard to solve. However,
projected gradient ascent method (PGAM), can be used to find the optimal
solutions. To apply the PGAM, first we define $\stackrel[i=1]{3}{\sum}\varpi_{u_{id}}\bar{R}_{u_{id}}+\varpi_{u_{iu}}\stackrel[i=1]{3}{\sum}\bar{R}_{u_{iu}}=f\left(\theta,\rho\right)$,
$\Phi=\left\{ \theta^{t}\in\mathbb{C}^{N\times1},\theta^{r}\in\mathbb{C}^{N\times1}\left|\left|\theta^{r}\right|=\left|\theta^{t}\right|=1\right.\right\} $,
and 

\noindent $Q=\left\{ \rho^{t}\in\mathbb{C}^{N\times1},\rho^{r}\in\mathbb{C}^{N\times1},\right.$ 

\noindent $\left.\left(\rho_{i}^{t}\right)+\left(\rho_{i}^{r}\right)=1,\rho_{i}^{t}\geq0,\rho_{i}^{r}\geq0\right\} $.
Then, we calculate the gradients of $f\left(\theta,\rho\right)$ with
respect to $\theta$, $\nabla_{\theta}f\left(\theta^{i},\rho^{i}\right)$,
and $\rho^{i}$, $\nabla_{\rho}f\left(\theta^{i},\rho^{i}\right)$.
Next we update the phases and amplitudes at each iteration using the
expressions, $\theta^{i+1}=\left(\theta^{i}+\nu_{i}\nabla_{\theta}f\left(\theta^{i},\rho^{i}\right)\right)$
and $\rho^{i+1}=\left(\rho^{i}+\vartheta_{i}\nabla_{\rho}f\left(\theta^{i},\rho^{i}\right)\right)$,
respectively, where $\nu_{i}$ and $\vartheta_{i}$ are the step sizes.
Then we project them onto $\Phi$ and $Q$. The overall steps of the
algorithm is summarized in Algorithm 2.

\begin{algorithm}[H]
Input: Set the maximum number of iterations $\mathscr{W}$ and a tolerance
$\mu>0$.

Initialize $\mathbf{\rho}^{(0)},\textrm{ and }\mathbf{\theta}^{(0)}$
, and set the step sizes.

for $i=0\textrm{ to }\mathscr{W}$ do

Evaluate: $f\left(\theta^{(i)},\rho^{(i)}\right)$, then $\nabla_{\theta}f\left(\theta^{(i)},\rho^{(i)}\right)$,
and $\nabla_{\rho}f\left(\theta^{(i)},\rho^{(i)}\right)$

Update: $\theta^{i+1}=\theta^{i}+\nu\nabla_{\theta}f\left(\theta^{i},\rho^{i}\right)$
and $\rho^{i+1}=\rho^{i}+\vartheta\nabla_{\rho}f\left(\theta^{i},\rho^{i}\right)$

Evaluate: $f\left(\theta^{(i+1)},\rho^{(i+1)}\right)$

Until Convergence: $f\left(\theta^{(i+1)},\rho^{(i+1)}\right)-f\left(\theta^{(i)},\rho^{(i)}\right)<\mu$.

\protect\caption{Optimization Algorithm.}
\end{algorithm}

\subsubsection{Sub-optimal Design}

\noindent In such systems the cell-edge users have the highest priority
in the STAR-RIS design. Therefore, the phase shifts can be aligned
to the cell-edge users' channels. The phase shifts can be presented
as, $\phi_{n}^{k}=-2\pi\frac{\iota}{\lambda}\left(c_{n}\upsilon_{b,u_{2k}}+f_{n}l_{b,u_{2k}}\right),k\in\left\{ t,r\right\} $
where $\upsilon_{u_{2k}}=\sin\varphi_{b,r}^{a}\sin\varphi_{b,r}^{e}-\sin\varphi_{r,u_{2k}}^{a}\sin\varphi_{r,u_{2k}}^{e},$
$l_{u_{2k}}=\cos\varphi_{b,r}^{e}-\cos\varphi_{r,u_{2k}}^{e}$, and
$\varphi_{i,j}^{a},\varphi_{i,j}^{e}$ are the azimuth and elevation
angles of arrival from node $i$ to node $j$, $\lambda$ is the wavelength,
$\iota$ is the elements spacing, and $c_{n}=\left(n-1\right)\textrm{mod}\sqrt{N}$,
$f_{n}=\frac{n-1}{\sqrt{N}}$. Also, the amplitudes $\rho_{n}^{r}$
and $\rho_{n}^{t}$ can be calculated using the ergodic rate expressions
based on the target data rates.

\subsection{Power Allocation}

The optimal power allocation scheme can be obtained by solving the
problem

\begin{eqnarray}
 & \underset{\mathbf{p}}{\max}\,\stackrel[i=1]{2}{\sum}f\left(\mathbf{p}\right)\nonumber \\
\left(C.1\right) & \textrm{s.t}\:\stackrel[i=1]{3}{\sum}P_{b_{i}}\leq P_{b},P_{b_{i}}>0,0<p_{u_{iu}}\leq p_{u_{m}}\nonumber \\
\left(C.2\right) & \bar{R}_{u_{id}}\geq\hat{R}_{d_{i}},\:\bar{R}_{u_{iu}}\geq\hat{R}_{u_{i}},\, i\in\left\{ 1,2,3\right\} \nonumber \\
\left(C.3\right) & R_{u_{kd}\rightarrow u_{id}}\geqslant R_{u_{id}},k>i\label{eq:30}
\end{eqnarray}

\noindent where $f\left(\mathbf{p}\right)=\stackrel[i=1]{3}{\sum}\omega_{u_{id}}\bar{R}_{u_{id}}+\stackrel[i=1]{3}{\sum}\omega_{u_{iu}}\bar{R}_{u_{iu}}$,
$\mathbf{p}=\left[p_{u_{1}},p_{u_{2}},p_{u_{3}},P_{b_{1}},P_{b_{2}},P_{b_{3}}\right]^{T}$,
$P_{b}=\stackrel[i=1]{3}{\sum}P_{b_{i}}$ and $p_{u_{m}}$ is the
maximum transmission power at a user. The optimal solution of the
problem in (\ref{eq:30}) can be obtained by using some complicated
techniques such as monotonic optimization, and block coordinate descent
(BCD) iterative algorithms which are described in \cite{powerallocation1,powerallocation2},
the details are omitted here due to the paper length limitation. 

\noindent As a simple sub-optimal solution, the minimum power required
for each user can be obtained by achieving the minimum data rate requirement
of each user, i.e., satisfying $(C.2)$ in (\ref{eq:30}) with equality,
$\bar{R}_{u_{id}}=\hat{R}_{d_{i}},\:\bar{R}_{u_{iu}}=\hat{R}_{u_{i}}$.
Thus, using the expressions in Theorems 1-6, the minimum required
power values can be obtained by solving all the equality equations
together.

\section{Numerical Results\label{sec:Numerical-Results}}

In this section simulation and numerical results are presented to
demonstrate the effectiveness of the proposed schemes and confirm
the accuracy of the analytical expressions. The radius of the cell-center
and cell-edge areas are assumed to be $R=50m$ and $R_{r}=30m$ respectively.
For simplicity, it is assumed that the nodes have same noise variance,
$\sigma^{2}$, and thus the DL transmit SNR is defined as $\bar{\gamma}=\frac{P_{b}}{\sigma^{2}}$,
and the maximum UL user power is $p_{u_{m}}=0.1P_{b}$ \cite{RefMain}.
The Rician factors are 3, and the path-loss exponent is $m=2.7$.
In addition, number of the STAR-RIS elements is $N=10$, number of
the cell-center users is $K_{c}$=6, and the cell-edge users $K_{e}=3$,
the self interference parameters are $\beta=0.001,\lambda=0.1$ and
the SIC error factor is $\Xi=0.1$.

\begin{figure}
\noindent \begin{centering}
\subfloat[\label{fig:1a}DL rates versus transmit SNR,$\bar{\gamma}$, for random
and optimal phase shifts. ]{\noindent \begin{centering}
\includegraphics[scale=0.5]{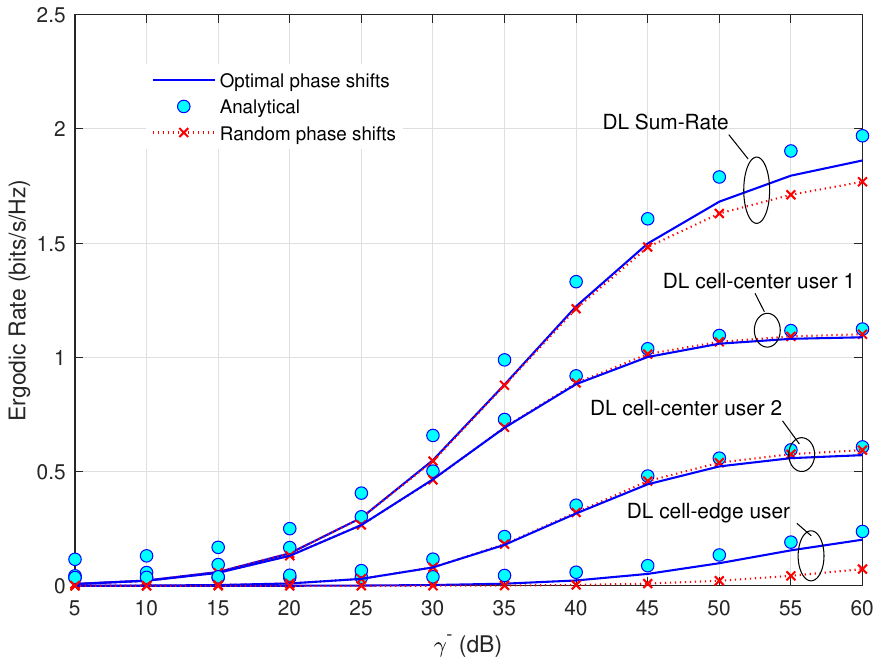}
\par\end{centering}

}
\par\end{centering}

\noindent \begin{centering}
\subfloat[\label{fig:1b}UL rates versus transmit SNR,$\bar{\gamma}$, for random
and optimal phase shifts.]{\noindent \begin{centering}
\includegraphics[scale=0.45]{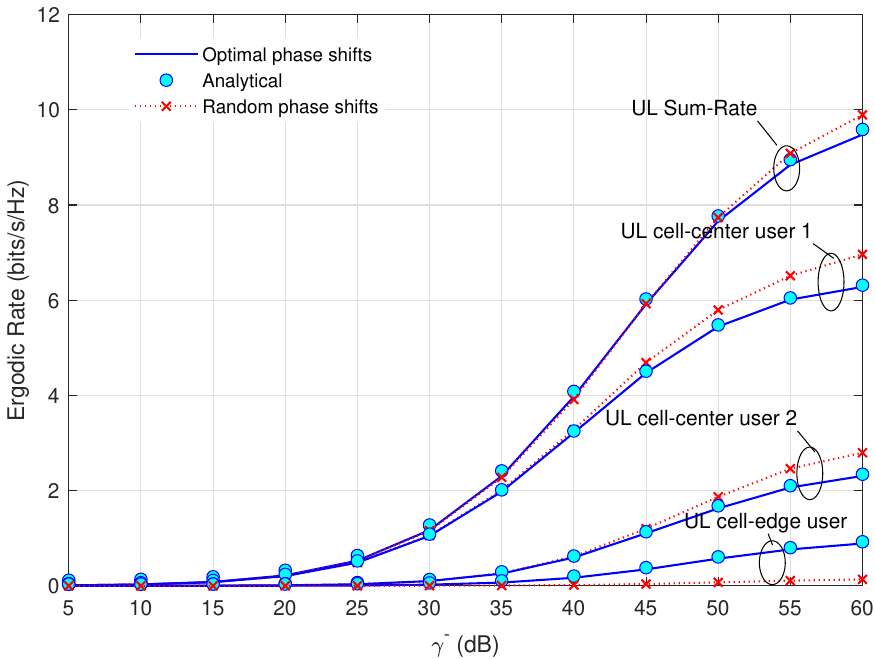}
\par\end{centering}

}
\par\end{centering}

\protect\caption{\label{fig:1}Ergodic rates versus transmit SNR,$\bar{\gamma}$, for
different phase shifts.}
\end{figure}

In Fig. \ref{fig:1} we illustrate the ergodic rates of a DL cluster
and an UL cluster versus the transmit SNR, $\bar{\gamma}$. Fig. \ref{fig:1a}
shows the achievable DL ergodic rates using the optimal and random
phase shifts, while the UL ergodic rates with the optimal and random
phase shifts are presented in Fig. \ref{fig:1b}. Firstly, it is clear
from these results that the DL and UL rates enhance with increasing
the transmit SNR. In addition, the cell-edge users achieve higher
data rates in the optimal phase shifts case than that in the random
phase shifts case. Interestingly enough, the performance of the UL
cell-center users improves in the random phase shifts case, due to
reducing the interference power caused by the UL cell-edge users. 

\begin{figure}
\noindent \begin{centering}
\subfloat[\label{fig:2a}DL rates versus transmit SNR$\bar{\gamma}$, with ideal
SIC $\Xi=0$. ]{\noindent \begin{centering}
\includegraphics[scale=0.5]{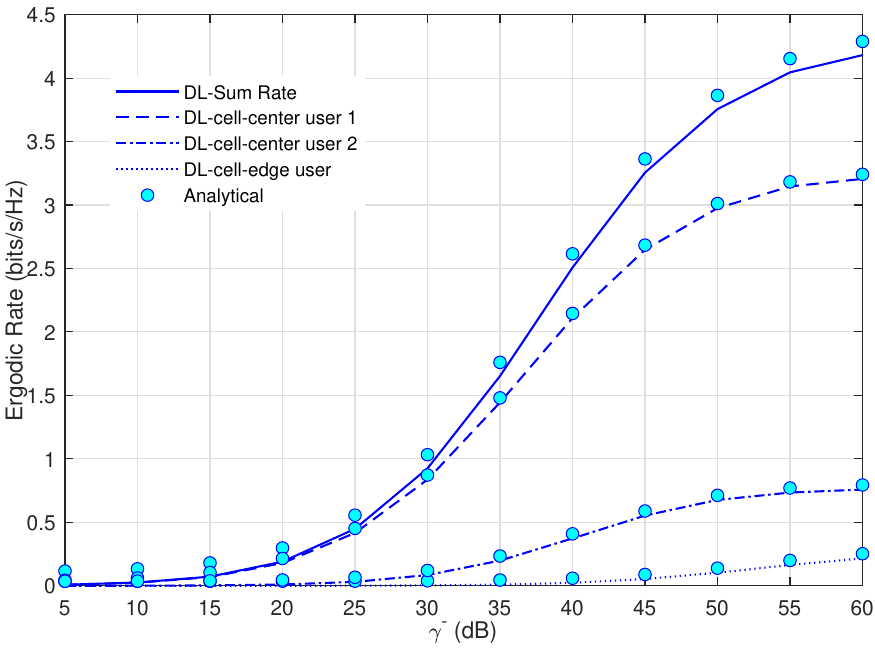}
\par\end{centering}

}
\par\end{centering}

\noindent \begin{centering}
\subfloat[\label{fig:2b}UL rates versus transmit SNR$\bar{\gamma}$, with ideal
SIC $\Xi=0$.]{\noindent \begin{centering}
\includegraphics[scale=0.5]{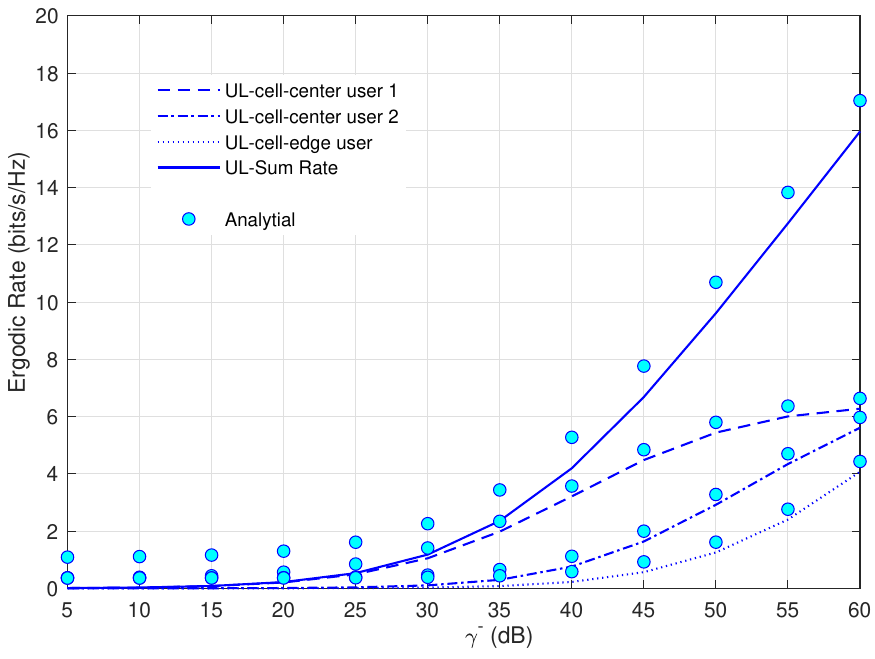}
\par\end{centering}

}
\par\end{centering}

\noindent \begin{centering}
\subfloat[\label{fig:2c}UL rates versus the transmit SNR, $\bar{\gamma}$,
when $\beta=1,\lambda=0.4$. ]{\noindent \centering{}\includegraphics[scale=0.5]{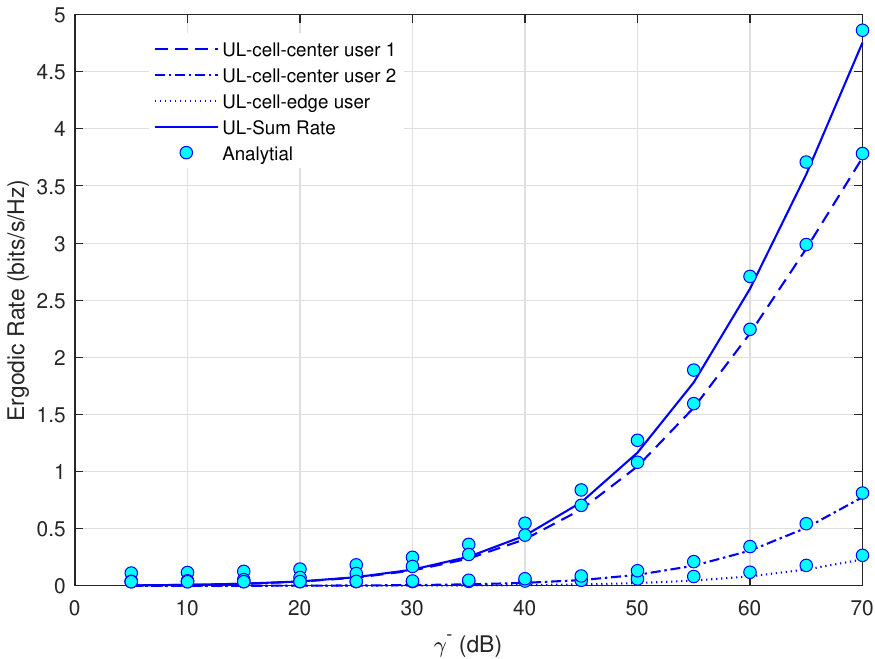}}
\par\end{centering}

\protect\caption{\label{fig:2}Ergodic rates versus transmit SNR,$\bar{\gamma}$, with
different values of the SIC error factor and self interference.}
\end{figure}

To show the impact of the system impairments on the users' performance,
in Fig. \ref{fig:2} we plot the DL and UL ergodic rates versus the
transmit SNR, $\bar{\gamma}$, for ideal SIC and high self interference
at the FD-BS. Fig. \ref{fig:2a} and Fig. \ref{fig:2b} depict the
DL and UL achievable rates with ideal SIC $\Xi=0$, respectively,
while Fig. \ref{fig:2c} presents the UL ergodic rates when the variance
of the residual self interference at the BS is high with, $\beta=1,\lambda=0.4$.
It can be observed from Figs. \ref{fig:2a} and \ref{fig:1a} that
the imperfect SIC results in deteriorating the achievable rates of
the DL cell-center users as their performance relies on the SIC detection
scheme. In addition, from Fig. \ref{fig:2b} and Fig. \ref{fig:1b},
we can see that, the performance of the UL cell-center user 2 and
the UL cell-edge user enhance greatly with reducing the SIC error
factor. Also, comparing Fig. \ref{fig:2c} and Fig. \ref{fig:1b},
we can observe that, high residual self interference at the BS causes
deep degradation of the UL users' performance. 

\begin{figure}
\noindent \begin{centering}
\subfloat[\label{fig:4a}DL sum-rates of user-clustering and user-paring schemes
versus the transmit SNR, $\bar{\gamma}$.]{\noindent \begin{centering}
\includegraphics[scale=0.5]{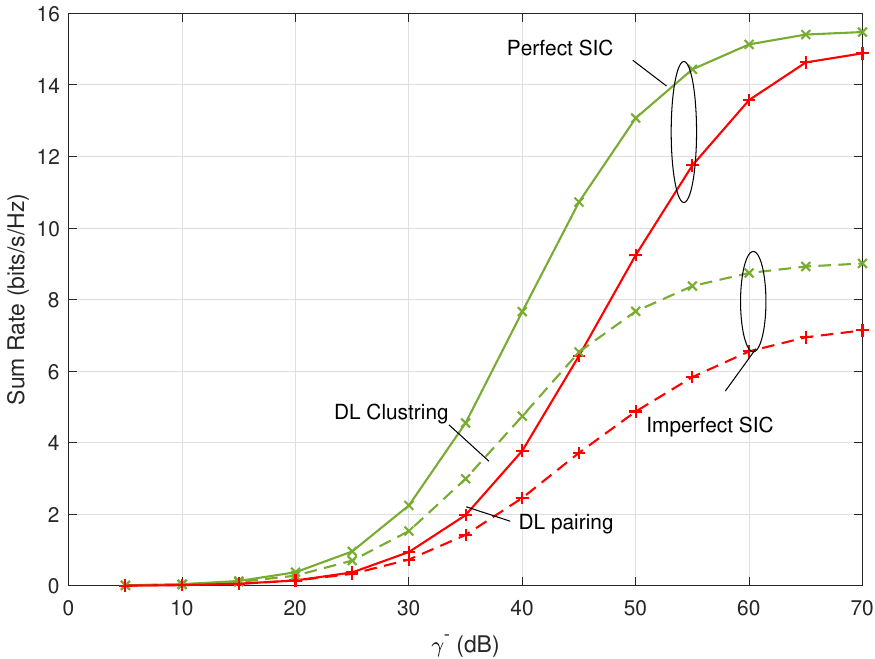}
\par\end{centering}

}
\par\end{centering}

\noindent \begin{centering}
\subfloat[\label{fig:4b}UL sum-rates of user-clustering and user-paring schemes
versus the transmit SNR, $\bar{\gamma}$.]{\noindent \begin{centering}
\includegraphics[scale=0.5]{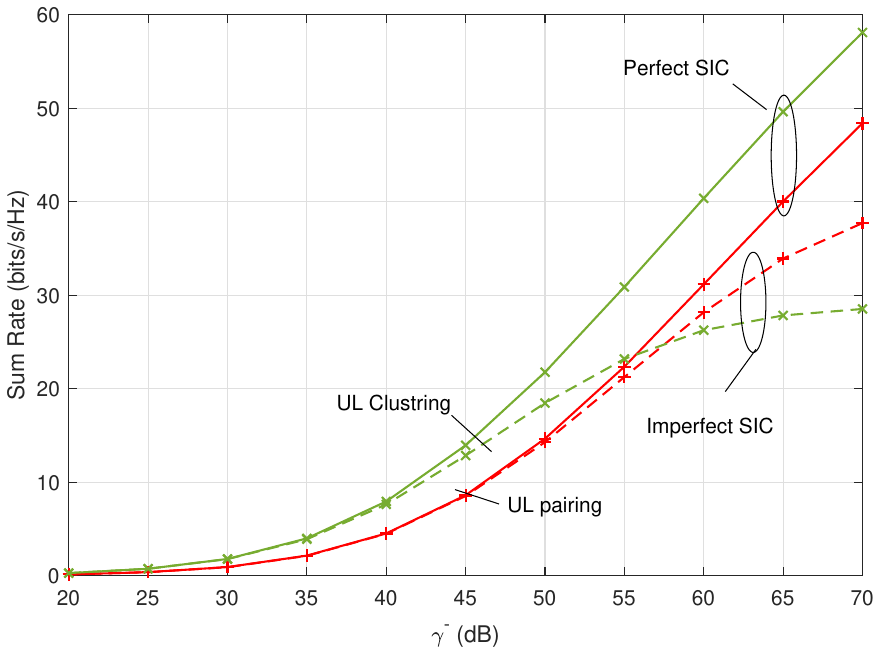}
\par\end{centering}

}
\par\end{centering}

\protect\caption{\label{fig:4}Sum-rates of user-clustering and user-paring schemes
versus the transmit SNR, $\bar{\gamma}$, for perfect and imperfect
SIC.}

\end{figure}

To compare NOMA user clustering scheme with NOMA pairing scheme considered
in \cite{part1}, we plot in Fig. \ref{fig:4} the DL and UL sum rates
of the two schemes versus the transmit SNR, $\bar{\gamma}$, for perfect
and imperfect SIC, $\Xi=0\textrm{ and }\Xi=0.1$,e.g., $0\%$ and
$10\%$ of the power remains as interference. The power in these results
is allocated between the users to achieve the target data rates of
the cell-edge users, and the remaining power is divided between the
cell-center users to enhance the total sum rate. It is worth mentioning
that, in NOMA user clustering scheme there are 3 different DL and
UL clusters, while in NOMA pairing scheme, there are 5 different DL
and UL clusters. From Fig. \ref{fig:4a}, in DL mode NOMA user clustering
scheme achieves higher sum rates than NOMA pairing scheme. On the
other hand, from Fig. \ref{fig:4b} in UL mode NOMA user clustering
scheme outperforms NOMA paring scheme in the perfect SIC case, and
at low SNR values in the imperfect SIC case. However, in a high SNR
regime NOMA pairing scheme can achieve higher sum-rates.  The sum-rate
of a cluster is controlled by the performance of the cell-center users.
In DL increasing SIC error hurts the performance of the cell-center
users in the two schemes, while in UL mode increasing SIC error degrades
the performance of the cell-edge users which can be controlled by
the power allocation scheme.  

\begin{figure}
\noindent \begin{centering}
\subfloat[\label{fig:5a}DL rates versus number of the STAR-RIS elements $N$. ]{\noindent \begin{centering}
\includegraphics[scale=0.5]{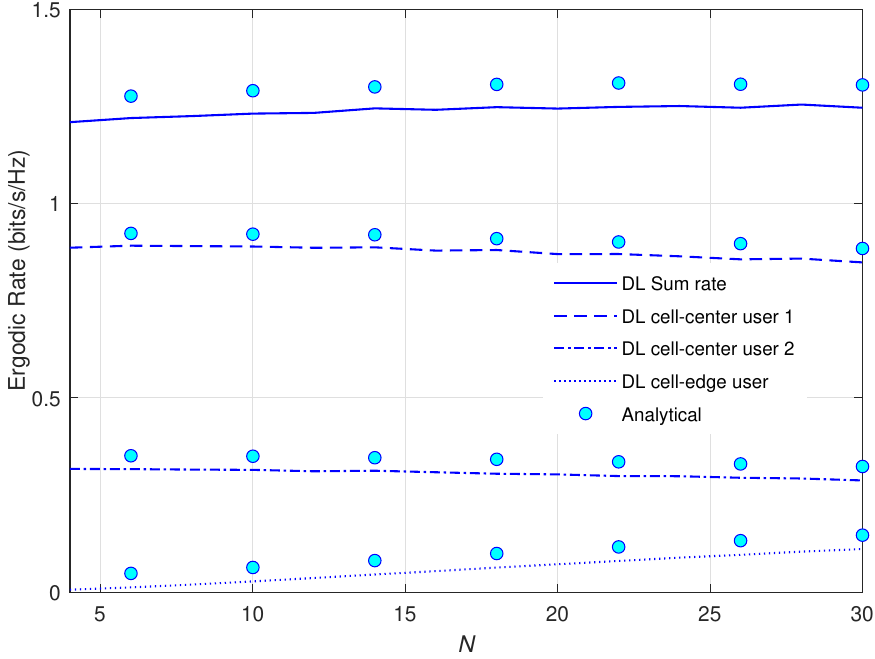}
\par\end{centering}

}
\par\end{centering}

\noindent \begin{centering}
\subfloat[\label{fig:5b}UL rates versus number of the STAR-RIS elements $N$. ]{\noindent \begin{centering}
\includegraphics[scale=0.5]{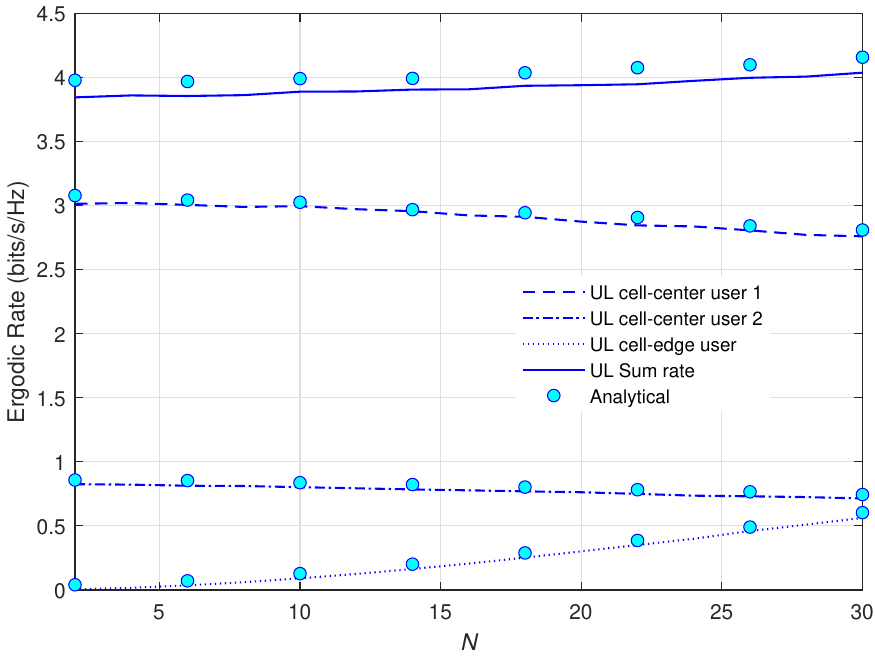}
\par\end{centering}

}
\par\end{centering}

\protect\caption{\label{fig:5}Ergodic rates versus number of the STAR-RIS elements
$N$.}

\end{figure}

Fig. \ref{fig:5}, illustrates the ergodic rates versus number of
the STAR-RIS elements $N$ when $\bar{\gamma}=40\textrm{ dB}$. Fig.
\ref{fig:5a} shows that the performance of the DL cell-edge user
is enhanced with increasing number of the STAR-RIS elements, $N$,
while the cell-center users performance degrades with increasing $N$.
This is because increasing number of the elements leads to increase
the interference power caused by the UL cell-edge user. On the other
hand, Fig. \ref{fig:4b} demonstrates that the performance of the
UL cell-edge user is enhanced greatly with increasing $N$, which
is not the case for the UL cell-center users where adding more elements
results in increasing the interference power caused by the UL cell-edge
user.

\section{Conclusions \label{sec:CONCLUSIONs} }

This work considered a STAR-RIS-assisted FD NOMA communication system,
where the STAR-RIS is implemented at the cell-edge region to assist
the cell-edge users. Firstly, new user clustering schemes for DL and
UL transmissions were presented. Then, closed-form analytical expressions
of the DL and UL ergodic sum rates have been derived. In addition,
the optimal amplitudes and phase-shifts of the STAR-RIS elements that
maximize the total sum-rate were obtained. Moreover, a power allocation
scheme of the DL and UL communications was studied. The results in
this work showed that increasing the transmitted SNR always improves
the achievable rates, and the performance of the cell-edge users can
be enhanced by using a large number of STAR-RIS elements. Furthermore,
imperfect SIC degrades the achievable rates of the DL cell-center
users and the UL cell-edge user, while imperfect SIS reduces the performance
of the UL users significantly. In addition, NOMA user clustering scheme
can achieve a higher sum rate than NOMA pairing scheme in DL mode,
while in UL mode NOMA user clustering scheme outperforms NOMA paring
scheme in the perfect SIC case, and at low SNR values in the imperfect
SIC case.

\section*{Appendix A}

By using Jensen inequality, the ergodic rate can be approximated by \cite{RISme,oldme,oldme2}

\[
\mathcal{E}\left[R_{u_{1d}}\right]\thickapprox\log_{2}\left(1+\right.
\]

\begin{equation}
\left.\frac{\alpha_{1}P_{b}\mathcal{E}\left\{ A_{u_{1d}}\right\} }{\Xi P_{b}\left(\alpha_{2}+\alpha_{3}\right)\mathcal{E}\left\{ A_{u_{1d}}\right\} +\mathcal{E}\left\{ B_{u_{1d}}\right\} +\mathcal{E}\left\{ C_{u_{1d}}\right\} +\mathcal{E}\left\{ D_{u_{1d}}\right\} +\sigma_{u_{1d}}^{2}}\right)
\end{equation}

1- The first term, $\mathcal{E}\left\{ A_{u_{1d}}\right\} $ can be
calculated as

\begin{equation}
\mathcal{E}\left\{ l_{b,u_{1d}}^{-m}\left|h_{b,u_{1d}}\right|^{2}\right\} =\mathcal{E}\left\{ l_{b,u_{1d}}^{-m}\right\} =x_{1_{u_{1d}}}
\end{equation}

The PDF of the users at radius $r$ relative to the BS is $f_{d}\left(r\right)=\frac{2r}{R^{2}},\quad0\leq r\leq R$.
Thus, the density of the $kth$ user order can be expressed as

\begin{equation}
f_{d_{k}}\left(r\right)=\frac{K_{cd}!}{\left(k-1\right)!\left(K_{cd}-1\right)!}\frac{2\left(r\right)}{\left(R\right)^{2}}\left(\frac{r^{2}}{R^{2}}\right)^{k-1}\left(1-\frac{r^{2}}{R^{2}}\right)^{K_{cd}-k}
\end{equation}
Then, we can find the average by 

\[
x_{1_{u_{1d}}}=\frac{2K_{cd}!}{\left(k_{cd_{1}}-1\right)!\left(K_{cd}-1\right)!}\stackrel[0]{R}{\int}\left(1+r_{u_{1d}}\right)^{-m}
\]

\begin{equation}
\times\frac{r}{\left(R\right)^{2}}\left(\frac{r^{2}}{\left(R\right)^{2}}\right)^{k_{cd_{1}}-1}\left(1-\frac{r^{2}}{R^{2}}\right)^{K_{cd}-k_{cd_{1}}}dr_{u_{1d}}\label{eq:34}
\end{equation}

\noindent The solution of (\ref{eq:34}) is presented in (\ref{eq:5}).

2- The second term, $\mathcal{E}\left\{ B_{u_{1d}}\right\} $ can
be calculated as

\begin{equation}
\mathcal{E}\left\{ \left|\sqrt{l_{u_{1d},u_{1u}}^{-m}}h_{u_{1d},u_{1u}}\right|^{2}\right\} =\mathcal{E}\left\{ l_{u_{1d},u_{1u}}^{-m}\right\} =y_{1_{u_{1d}}}\label{eq:35}
\end{equation}

\noindent Considering the distribution of the distance between two
random points inside a circle in \cite{booknew}, we can write 

\[
y_{1_{u_{1d}}}=\stackrel[_{0}]{2R}{\int}\left(1+r_{u_{1d},u_{1u}}\right)^{-m}\frac{4r_{u_{1d},u_{1u}}}{\pi R^{2}}
\]

\begin{equation}
\times\left(\cos^{-1}\left(\frac{r_{u_{1d},u_{1u}}}{2R}\right)-\frac{r_{u_{1d},u_{1u}}}{2R}\left(\sqrt{1-\frac{r_{u_{1d},u_{1u}}^{2}}{4R^{2}}}\right)\right)dr_{u_{1d},u_{1u}}\label{eq:36}
\end{equation}

\noindent The solution of (\ref{eq:36}) is presented in (\ref{eq:7}).

3- The third term, $\mathcal{E}\left\{ C_{u_{1d}}\right\} $ can be
calculated by following similar steps as in (\ref{eq:35}) and (\ref{eq:36}).

4- The last term $\mathcal{E}\left\{ D_{u_{1d}}\right\} $ can be
evaluated as 

\[
\mathcal{E}\left\{ p_{u_{3u}}l_{r,u3u}^{-m}l_{r,u1d}^{-m}\left|g_{r,u_{1d}}\Theta_{t}g_{r,u_{3u}}\right|^{2}\right\} =
\]

\begin{equation}
p_{u_{3u}}\mathcal{E}\left\{ l_{r,u3u}^{-m}\right\} \mathcal{E}\left\{ l_{r,u1d}^{-m}\right\} \mathcal{E}\left\{ \left|g_{r,u_{1d}}\Theta_{t}g_{r,u_{3u}}\right|^{2}\right\} 
\end{equation}

\noindent The PDF of the weak user location at radius $r_{r}$ relative
to the RIS is $f_{d}\left(r_{r}\right)=\frac{2r_{r}}{R_{r}^{2}},\quad0\leq r_{r}\leq R_{r}$.
Thus, we can calculate the first average as,

\[
\mathcal{E}\left\{ l_{r,u3u}^{-m}\right\} =\frac{2K_{eu}!}{\left(k-1\right)!\left(K_{eu}-k\right)!}\stackrel[0]{R_{r}}{\int}\left(1+r_{r,u3u}\right)^{-m}
\]

\begin{equation}
\times\frac{r}{\left(R_{r}\right)^{2}}\left(\frac{r^{2}}{\left(R_{r}\right)^{2}}\right)^{k-1}\left(1-\frac{r^{2}}{R_{r}^{2}}\right)^{K_{eu}-k}dr_{r,u3u}
\end{equation}

\noindent which can be found as 

\[
y_{2_{u_{1d}}}=\left(2a_{1_{u3u}}\Gamma\left(k_{eu_{3}}\right)\mathcal{H}\left(\left\{ k_{eu_{3}},\frac{1+m}{2},\frac{m}{2}\right\} ,\left\{ \frac{1}{2},1+K_{eu}\right\} ,R_{r}^{2}\right)\right.
\]

\begin{equation}
\left.-mR_{r}a_{1_{u3u}}\Gamma\left(\frac{1}{2}+k_{eu_{3}}\right)\mathcal{H}\left(\left\{ \frac{1}{2}+k_{eu_{3}},\frac{1+m}{2},\frac{2+m}{2}\right\} ,\left\{ \frac{3}{2},\frac{3}{2}+K_{eu}\right\} ,R_{r}^{2}\right)\right)\label{eq:39}
\end{equation}

\noindent By invoking the distribution of the distance between a random
point inside a circle and a fixed point outside the circle provided
in \cite{booknew}, we can write 

\[
\mathcal{E}\left\{ l_{r,u_{1d}}^{-m}\right\} =\stackrel[r_{1}]{r_{1}+2R}{\int}r_{r,u1d}^{-m}\frac{2r_{r,u_{1d}}}{\pi R^{2}}\cos^{-1}\left(\frac{1}{r_{r,u_{1d}}}\right.
\]

\begin{equation}
\left.\left(r_{1}+\frac{\left(r_{r,u_{1d}}^{2}-r_{1}^{2}\right)}{2\left(R+r_{1}\right)}\right)\right)dr_{r,u_{1d}}\label{eq:sup7}
\end{equation}

\noindent where $r_{1}=d_{b,r}-R$ is the distance from RIS to the
circle boundary, $r_{1}\leq r_{r,u_{1d}}\leq r_{1}+2R$. Applying
Gaussian Quadrature rules we can get

\[
\mathcal{E}\left\{ l_{r,u1d}^{-m}\right\} =\stackrel[j=1]{C}{\sum}\textrm{H}_{j}\,\left(1+\left(R\, r_{j}+R\right)\right)^{-m}\frac{2\left(R\, r_{j}+R\right)}{\pi R^{2}}
\]

\begin{equation}
\times\cos^{-1}\left(\frac{1}{\left(R\, r_{j}+R\right)}\left(r_{1}+\frac{\left(\left(R\, r_{j}+R\right)^{2}-r_{1}^{2}\right)}{2\left(R+r_{1}\right)}\right)\right)
\end{equation}
After removing the zero expectation terms, the last expectation can
be written as

\[
\mathscr{E}\left\{ \left|\mathbf{g}_{_{r,u1d}}\Theta\mathbf{g}_{r,u_{3u}}\right|^{2}\right\} =\frac{\kappa_{r,u1d}}{\kappa_{r,u1d}+1}\frac{\kappa_{r,u_{3u}}}{\kappa_{r,u_{3u}}+1}\mathscr{E}\left|\mathbf{\bar{g}}_{_{r,u1d}}\Theta\bar{\mathbf{g}}_{r,u_{3u}}\right|^{2}+\frac{\kappa_{r,u1d}}{\kappa_{r,u1d}+1}\frac{1}{\kappa_{r,u_{3u}}+1}\mathscr{E}\left|\mathbf{\bar{g}}_{_{r,u1d}}\Theta\tilde{\mathbf{g}}_{r,u_{3u}}\right|^{2}
\]

\begin{equation}
+\frac{\kappa_{r,u_{3u}}}{\kappa_{r,u_{3u}}+1}\frac{1}{\kappa_{r,u1d}+1}\mathscr{E}\left|\mathbf{\tilde{g}}_{_{r,u1d}}\Theta\bar{\mathbf{g}}_{r,u_{3u}}\right|^{2}+\frac{1}{\kappa_{r,u1d}+1}\frac{1}{\kappa_{r,u_{3u}}+1}\mathscr{E}\left|\mathbf{\tilde{g}}_{_{r,u1d}}\Theta\mathbf{\tilde{g}}_{r,u_{3u}}\right|^{2}\label{eq:sup10}
\end{equation}

Now, the first term in (\ref{eq:sup10}) is 

\[
\mathscr{E}\left|\mathbf{\bar{g}}_{_{r,u1d}}\Theta\bar{\mathbf{g}}_{r,u_{3u}}\right|^{2}=
\]

\begin{equation}
=\left|\stackrel[n=1]{N}{\sum}a_{N,n}\left(\psi_{r,u_{3u}}^{a},\psi_{r,u_{3u}}^{e}\right)\rho_{n}^{k}e^{j\phi_{n}^{k}}a_{N,n}\left(\psi_{r,u_{1d}}^{a},\psi_{r,u_{1d}}^{e}\right)\right|^{2}=\xi_{1}
\end{equation}

Similarly, the second term,

\[
\mathscr{E}\left|\mathbf{\bar{g}}_{_{r,u1d}}\Theta\tilde{\mathbf{g}}_{r,u_{3u}}\right|^{2}=\stackrel[n=1]{N}{\sum}\left|\rho_{n}^{k}\right|^{2}+
\]

\[
\mathscr{E}\left\{ \stackrel[n_{1}=1]{N}{\sum}\stackrel[n_{2}\neq n_{1}]{N}{\sum}\left(a_{Nn_{1}}\left(\psi_{r,u_{1d}}^{a},\psi_{r,u_{1d}}^{e}\right)\rho_{n_{1}}^{k}e^{j\phi_{n_{1}}^{k}}\left[\tilde{\mathbf{g}}_{r,u_{3u}}\right]_{n_{1}}\right)\right.
\]

\begin{equation}
\left.\left(a_{Nn_{2}}\left(\psi_{r,u_{1d}}^{a},\psi_{r,u_{1d}}^{e}\right)\rho_{n_{2}}^{k}e^{j\phi_{n_{2}}^{k}}\left[\tilde{\mathbf{g}}_{r,u_{3u}}\right]_{n_{2}}\right)^{H}\right\} =\stackrel[n=1]{N}{\sum}\left|\rho_{n}^{k}\right|^{2}
\end{equation}

\noindent The other terms,

\[
\mathscr{E}\left|\mathbf{\tilde{g}}_{_{r,u1d}}\Theta\bar{\mathbf{g}}_{r,u_{3u}}\right|^{2}=\stackrel[n=1]{N}{\sum}\left|\rho_{n}^{k}\right|^{2}+
\]

\[
\mathscr{E}\left\{ \stackrel[n_{1}=1]{N}{\sum}\stackrel[n_{2}\neq n_{1}]{N}{\sum}\left(\left[\mathbf{\tilde{g}}_{_{r,u1d}}\right]_{n_{1}}\rho_{n_{1}}^{k}e^{j\phi_{n_{1}}^{k}}a_{Nn1}\left(\psi_{r,u_{3u}}^{a},\psi_{r,u_{3u}}^{e}\right)\right)\right.
\]

\begin{equation}
\left.\left(\left[\mathbf{\tilde{g}}_{_{r,u1d}}\right]_{n_{2}}\rho_{n_{1}}^{k}e^{j\phi_{n_{1}}^{k}}a_{Nn_{2}}\left(\psi_{r,u_{3u}}^{a},\psi_{r,u_{3u}}^{e}\right)\right)^{H}\right\} =\stackrel[n=1]{N}{\sum}\left|\rho_{n}^{k}\right|^{2}
\end{equation}

\noindent and

\[
\mathscr{E}\left|\mathbf{\tilde{g}}_{_{r,u1d}}\Theta\mathbf{\tilde{g}}_{r,u_{3u}}\right|^{2}=
\]

\begin{equation}
\mathscr{E}\left|\stackrel[n=1]{N}{\sum}\left[\mathbf{\tilde{g}}_{_{r,u1d}}\right]_{n}\rho_{n}^{k}e^{j\phi_{n}^{k}}\left[\mathbf{\tilde{g}}_{r,u_{3u}}\right]_{n}\right|^{2}=\stackrel[n=1]{N}{\sum}\left|\rho_{n}^{k}\right|^{2}
\end{equation}

Now, we are ready to write the average as 

\[
\mathscr{E}\left\{ \left|\mathbf{g}_{r,u_{1d}}\Theta\mathbf{g}_{r,u_{3u}}\right|^{2}\right\} =\frac{\kappa_{r,u_{1d}}}{\kappa_{r,u_{1d}}+1}\frac{\kappa_{r,u_{3u}}}{\kappa_{r,u_{3u}}+1}\xi_{1}
\]

\[
+\frac{\kappa_{r,u_{1d}}}{\kappa_{r,u_{1d}}+1}\frac{\stackrel[n=1]{N}{\sum}\left|\rho_{n}^{k}\right|^{2}}{\kappa_{r,u_{3u}}+1}+\frac{\kappa_{r,u_{3u}}}{\kappa_{r,u_{3u}}+1}\frac{\stackrel[n=1]{N}{\sum}\left|\rho_{n}^{k}\right|^{2}}{\kappa_{r,u_{1d}}+1}
\]

\begin{equation}
+\frac{1}{\kappa_{r,u_{1d}}+1}\frac{1}{\kappa_{r,u_{3u}}+1}\stackrel[n=1]{N}{\sum}\left|\rho_{n}^{k}\right|^{2}
\end{equation}

\section*{Appendix B}

By using Jensen inequality, the ergodic rate can be approximated by

\[
\mathcal{E}\left[R_{u_{2d}}\right]\thickapprox\log_{2}\left(1+\right.
\]

\begin{equation}
\left.\frac{\alpha_{2}P_{b}\mathcal{E}\left\{ A_{u_{2d}}\right\} }{P_{b}\mathcal{E}\left\{ A_{u_{2d}}\right\} \left(\Xi\alpha_{3}+\alpha_{1}\right)+\mathcal{E}\left\{ B_{u_{2d}}\right\} +\mathcal{E}\left\{ C_{u_{2d}}\right\} +\mathcal{E}\left\{ D_{u_{2d}}\right\} +\sigma_{u_{2d}}^{2}}\right)
\end{equation}

1- The first term, $\mathcal{E}\left\{ A_{u_{2d}}\right\} $ can be
calculated as

\begin{equation}
\mathcal{E}\left\{ \left|\sqrt{l_{b,u_{2d}}^{-m}}h_{u_{2d}}\right|^{2}\right\} =\mathcal{E}\left\{ l_{b,u_{2d}}^{-m}\right\} 
\end{equation}

which can be found by, 

\begin{equation}
\mathcal{E}\left\{ l_{b,u_{2d}}^{-m}\right\} =\frac{2K_{cd}!}{\left(k_{cd_{2}}-1\right)!\left(K_{cd}-1\right)!}\stackrel[0]{R}{\int}\left(1+r_{u_{2d}}\right)^{-m}\frac{\left(r\right)}{\left(R\right)^{2}}\left(\frac{r^{2}}{R^{2}}\right)^{k_{cd_{2}}-1}\left(1-\frac{r^{2}}{R^{2}}\right)^{K_{cd}-k_{cd_{2}}}dr_{u_{2d}}
\end{equation}

\[
\mathcal{E}\left\{ l_{b,u_{2d}}^{-m}\right\} =2a_{1_{u2d}}\Gamma\left(k_{cd_{2}}\right)\mathcal{H}\left(\left\{ k_{cd_{2}},\frac{1+m}{2},\frac{m}{2}\right\} ,\left\{ \frac{1}{2},1+K_{cd}\right\} ,R^{2}\right)-
\]

\begin{equation}
mRa_{1_{u2d}}\Gamma\left(\frac{1}{2}+k_{cd_{2}}\right)\mathcal{H}\left(\left\{ \frac{1}{2}+k_{cd_{2}},\frac{1+m}{2},\frac{2+m}{2}\right\} ,\left\{ \frac{3}{2},\frac{3}{2}+K_{cd}\right\} ,R^{2}\right)\label{eq:8-2}
\end{equation}

The derivation of $\mathcal{E}\left\{ B_{u_{2d}}\right\} ,\mathcal{E}\left\{ C_{u_{2d}}\right\} ,\mathcal{E}\left\{ D_{u_{2d}}\right\} $
can be obtained as in Appendix A.

\section*{Appendix C}

By using Jensen inequality, the ergodic rate can be approximated by

\[
\mathcal{E}\left[R_{u_{3d}}\right]\thickapprox\log_{2}\left(1+\right.
\]

\begin{equation}
\left.\frac{\mathcal{E}\left\{ A_{u_{3d}}\right\} }{P_{b_{1,2}}\mathcal{E}\left\{ A_{u_{3d}}\right\} +\mathcal{E}\left\{ B_{u_{3d}}\right\} +\mathcal{E}\left\{ C_{u_{3d}}\right\} +\mathcal{E}\left\{ D_{u_{3d}}\right\} +\sigma_{u_{3d}}^{2}}\right)
\end{equation}

1-The first term, $\mathcal{E}\left\{ A_{u_{3d}}\right\} $ can be
calculated as

\begin{equation}
\mathcal{E}\left\{ l_{b,r}^{-m}l_{r,u3d}^{-m}\left|\mathbf{g}_{_{r,u3d}}\Theta\mathbf{g}_{b,r}\right|^{2}\right\} =l_{b,r}^{-m}\mathcal{E}\left\{ l_{r,u3d}^{-m}\right\} \mathcal{E}\left\{ \left|\mathbf{g}_{_{r,u3d}}\Theta\mathbf{g}_{b,r}\right|^{2}\right\} 
\end{equation}

where 

\begin{equation}
\mathcal{E}\left\{ l_{r,u3d}^{-m}\right\} =\frac{2K_{ed}!}{\left(k{}_{ed_{3}}-1\right)!\left(K_{ed}-1\right)!}\stackrel[0]{R_{r}}{\int}r_{r,u3d}^{-m}\frac{r}{\left(R_{r}\right)^{2}}\left(\frac{r^{2}}{\left(R_{r}\right)^{2}}\right)^{k'_{e}-1}\left(1-\frac{r^{2}}{R_{r}^{2}}\right)^{K_{ed}-k{}_{ed_{3}}}dr_{r,u3d}
\end{equation}

\[
\mathcal{E}\left\{ l_{r,u3d}^{-m}\right\} =\frac{\sqrt{\pi}\Gamma\left(1-k{}_{ed}+K_{ed}\right)2K_{ed}!}{4\left(k{}_{ed_{3}}-1\right)!\left(K_{ed}-k{}_{ed_{3}}\right)!}\left(2\Gamma\left(k{}_{ed_{3}}\right)\mathcal{H}\left(\left\{ k{}_{ed_{3}},\frac{1+m}{2},\frac{m}{2}\right\} ,\left\{ \frac{1}{2},1+K_{ed}\right\} ,R_{r}^{2}\right)\right.
\]

\begin{equation}
\left.-mR_{r}\Gamma\left(\frac{1}{2}+k{}_{ed_{3}}\right)\mathcal{H}\left(\left\{ \frac{1}{2}+k{}_{ed_{3}},\frac{1+m}{2},\frac{2+m}{2}\right\} ,\left\{ \frac{3}{2},\frac{3}{2}+K_{ed}\right\} ,R_{r}^{2}\right)\right)\label{eq:8-1-1}
\end{equation}

and

\[
\mathscr{E}\left\{ \left|\mathbf{g}_{_{r,u3d}}\Theta\mathbf{g}_{b,r}\right|^{2}\right\} =\frac{\kappa_{r,u3d}}{\kappa_{r,u3d}+1}\frac{\kappa_{b,r}}{\kappa_{b,r}+1}\xi_{3}+\frac{\kappa_{r,u3d}}{\kappa_{r,u3d}+1}\frac{1}{\kappa_{b,r}+1}\stackrel[n=1]{N}{\sum}\left|\rho_{n}^{k}\right|^{2}
\]

\begin{equation}
+\frac{\kappa_{b,r}}{\kappa_{b,r}+1}\frac{1}{\kappa_{r,u3d}+1}\stackrel[n=1]{N}{\sum}\left|\rho_{n}^{k}\right|^{2}+\frac{1}{\kappa_{r,u3d}+1}\frac{1}{\kappa_{b,r}+1}\stackrel[n=1]{N}{\sum}\left|\rho_{n}^{k}\right|^{2}
\end{equation}

\noindent where $\xi_{3}=\left|\mathbf{\bar{g}}_{_{r,u3d}}\Theta\bar{\mathbf{g}}_{b,r}\right|^{2}.$

2- The term, $\mathcal{E}\left\{ B_{u_{3d}}\right\} $ can be calculated
as

\begin{equation}
\mathcal{E}\left\{ l_{r,u3d}^{-m}l_{r,u1u}^{-m}\left|\mathbf{g}_{_{r,u3d}}\Theta_{k}\mathbf{g}_{u_{1u,r}}\right|^{2}\right\} =\mathcal{E}\left\{ l_{r,u3d}^{-m}\right\} \mathcal{E}\left\{ l_{r,u1u}^{-m}\right\} \mathcal{E}\left\{ \left|\mathbf{g}_{_{r,u3d}}\Theta_{k}\mathbf{g}_{u_{1u,r}}\right|^{2}\right\} 
\end{equation}

\begin{equation}
\mathcal{E}\left\{ l_{r,u1u}^{-m}\right\} =\stackrel[r_{1}]{r_{1}+2R}{\int}r_{r,u1u}^{-m}\frac{2r_{r,u1u}}{\pi R^{2}}\cos^{-1}\left(\frac{1}{r_{r,u1u}}\left(r_{1}+\frac{\left(r_{r,u1u}^{2}-r_{1}^{2}\right)}{2\left(R+r_{1}\right)}\right)\right)dr_{r,u1u}
\end{equation}

\noindent where $r_{1}\leq r_{r,u1u}\leq r_{1}+2R$, $r_{1}=d_{b,r}-R$
is the distance from RIS to the boundary. Applying Gaussian Quadrature
rules we can get,

\begin{equation}
\mathcal{E}\left\{ l_{r,u1u}^{-m}\right\} =\stackrel[j=1]{C}{\sum}\textrm{H}_{j}\,\left(1+\left(R\, r_{j}+R\right)\right)^{-m}\frac{2\left(R\, r_{j}+R\right)}{\pi R^{2}}\cos^{-1}\left(\frac{1}{\left(R\, r_{j}+R\right)}\left(r_{1}+\frac{\left(\left(R\, r_{j}+R\right)^{2}-r_{1}^{2}\right)}{2\left(R+r_{1}\right)}\right)\right)
\end{equation}

Following similar steps we can find 

\[
\mathcal{E}\left\{ l_{r,u3d}^{-m}\right\} =\frac{\sqrt{\pi}\Gamma\left(1-k{}_{ed_{3}}+K_{ed}\right)2K_{ed}!}{4\left(k{}_{ed_{3}}-1\right)!\left(K_{ed}-k{}_{ed_{3}}\right)!}\left(2\Gamma\left(k{}_{ed_{3}}\right)\mathcal{H}\left(\left\{ k{}_{ed_{3}},\frac{1+m}{2},\frac{m}{2}\right\} ,\left\{ \frac{1}{2},1+K_{ed}\right\} ,R_{r}^{2}\right)\right.
\]

\begin{equation}
\left.-mR_{r}\Gamma\left(\frac{1}{2}+k{}_{ed_{3}}\right)\mathcal{H}\left(\left\{ \frac{1}{2}+k{}_{ed_{3}},\frac{1+m}{2},\frac{2+m}{2}\right\} ,\left\{ \frac{3}{2},\frac{3}{2}+K_{ed}\right\} ,R_{r}^{2}\right)\right)
\end{equation}

and

\[
\mathscr{E}\left\{ \left|\mathbf{g}_{_{r,u3d}}\Theta_{k}\mathbf{g}_{u_{1u,r}}\right|^{2}\right\} =\frac{\kappa_{r,u3d}}{\kappa_{r,u3d}+1}\frac{\kappa_{u_{1u,r}}}{\kappa_{u_{1u,r}}+1}\xi_{4}+\frac{\kappa_{r,u3d}}{\kappa_{r,u3d}+1}\frac{1}{\kappa_{u_{1u,r}}+1}\stackrel[n=1]{N}{\sum}\left|\rho_{n}^{k}\right|^{2}
\]

\begin{equation}
+\frac{\kappa_{u_{1u,r}}}{\kappa_{u_{1u,r}}+1}\frac{1}{\kappa_{r,u3d}+1}\stackrel[n=1]{N}{\sum}\left|\rho_{n}^{k}\right|^{2}+\frac{1}{\kappa_{r,u3d}+1}\frac{1}{\kappa_{u_{1u,r}}+1}\stackrel[n=1]{N}{\sum}\left|\rho_{n}^{k}\right|^{2}
\end{equation}

\noindent where $\xi_{4}=\left|\mathbf{\bar{g}}_{_{r,u3d}}\Theta\bar{\mathbf{g}}_{u_{1u,r}}\right|^{2}.$

3- The term, $\mathcal{E}\left\{ C_{u_{3d}}\right\} $ can be calculated
by

\begin{equation}
\mathcal{E}\left\{ l_{r,u3d}^{-m}l_{r,u2u}^{-m}\left|\mathbf{g}_{_{r,u3d}}\Theta\mathbf{g}_{u_{2u,r}}\right|^{2}\right\} =\mathcal{E}\left\{ l_{r,u3d}^{-m}\right\} \mathcal{E}\left\{ l_{r,u2u}^{-m}\right\} \mathcal{E}\left\{ \left|\mathbf{g}_{_{r,u3d}}\Theta\mathbf{g}_{u_{2u,r}}\right|^{2}\right\} 
\end{equation}

The first expectation can be found as

\begin{equation}
\mathcal{E}\left\{ l_{r,u2u}^{-m}\right\} =\stackrel[r_{1}]{r_{1}+2R}{\int}r_{r,u2u}^{-m}\frac{2r_{r,u2u}}{\pi R^{2}}\cos^{-1}\left(\frac{1}{r_{r,u2u}}\left(r_{1}+\frac{\left(r_{r,u2u}^{2}-r_{1}^{2}\right)}{2\left(R+r_{1}\right)}\right)\right)dr_{r,u2u}
\end{equation}

\[
\mathcal{E}\left\{ l_{r,u2u}^{-m}\right\} =\stackrel[j=1]{C}{\sum}\textrm{H}_{j}\,\left(1+\left(R\, r_{j}+R\right)\right)^{-m}\frac{2\left(R\, r_{j}+R\right)}{\pi R^{2}}
\]

\begin{equation}
\times\cos^{-1}\left(\frac{1}{\left(R\, r_{j}+R\right)}\left(r_{1}+\frac{\left(\left(R\, r_{j}+R\right)^{2}-r_{1}^{2}\right)}{2\left(R+r_{1}\right)}\right)\right)
\end{equation}

From the previous derivations, we can write 

\[
\mathcal{E}\left\{ l_{r,u3d}^{-m}\right\} =\frac{\sqrt{\pi}\Gamma\left(1-k{}_{ed_{3}}+K_{ed}\right)2K_{ed}!}{4\left(k{}_{ed_{3}}-1\right)!\left(K_{ed}-k{}_{ed_{3}}\right)!}\left(2\Gamma\left(k{}_{ed_{3}}\right)\mathcal{H}\left(\left\{ k{}_{ed_{3}},\frac{1+m}{2},\frac{m}{2}\right\} ,\left\{ \frac{1}{2},1+K_{ed}\right\} ,R_{r}^{2}\right)\right.
\]

\begin{equation}
\left.-mR_{r}\Gamma\left(\frac{1}{2}+k{}_{ed_{3}}\right)\mathcal{H}\left(\left\{ \frac{1}{2}+k{}_{ed_{3}},\frac{1+m}{2},\frac{2+m}{2}\right\} ,\left\{ \frac{3}{2},\frac{3}{2}+K_{ed}\right\} ,R_{r}^{2}\right)\right)
\end{equation}

\[
\mathscr{E}\left\{ \left|\mathbf{g}_{_{r,u3d}}\Theta\mathbf{g}_{u_{2u,r}}\right|^{2}\right\} =\frac{\kappa_{r,u3d}}{\kappa_{r,u3d}+1}\frac{\kappa_{u_{2u,r}}}{\kappa_{u_{2u,r}}+1}\xi_{5}+\frac{\kappa_{r,u3d}}{\kappa_{r,u3d}+1}\frac{1}{\kappa_{u_{2u,r}}+1}\stackrel[n=1]{N}{\sum}\left|\rho_{n}^{k}\right|^{2}
\]

\begin{equation}
+\frac{\kappa_{u_{1u,r}}}{\kappa_{u_{1u,r}}+1}\frac{1}{\kappa_{u_{2u,r}}+1}\stackrel[n=1]{N}{\sum}\left|\rho_{n}^{k}\right|^{2}+\frac{1}{\kappa_{r,u3d}+1}\frac{1}{\kappa_{u_{2u,r}}+1}\stackrel[n=1]{N}{\sum}\left|\rho_{n}^{k}\right|^{2}
\end{equation}

\noindent where $\xi_{5}=\left|\mathbf{\bar{g}}_{_{r,u3d}}\Theta\bar{\mathbf{g}}_{u_{2u,r}}\right|^{2}.$

4- Similarly, the term, $\mathcal{E}\left\{ D_{u_{3d}}\right\} $
can be calculated as

\noindent 
\begin{equation}
\mathcal{E}\left\{ D_{u_{3d}}\right\} =p_{u_{3u}}\mathcal{E}\left\{ l_{r,u_{3d}}^{-m}\right\} \mathcal{E}\left\{ l_{r,u_{3u}}^{-m}\right\} \mathcal{E}\left\{ \left|\mathbf{g}_{r,u_{3d}}\Theta_{r}\mathbf{g}_{r,u_{3u}}\right|^{2}\right\} 
\end{equation}

where

\[
\mathcal{E}\left\{ l_{r,u3d}^{-m}\right\} =\frac{\sqrt{\pi}\Gamma\left(1-k{}_{ed_{3}}+K_{ed}\right)2K_{ed}!}{4\left(k{}_{ed_{3}}-1\right)!\left(K_{ed}-k{}_{ed_{3}}\right)!}\left(2\Gamma\left(k{}_{ed_{3}}\right)\mathcal{H}\left(\left\{ k{}_{ed_{3}},\frac{1+m}{2},\frac{m}{2}\right\} ,\left\{ \frac{1}{2},1+K_{ed}\right\} ,R_{r}^{2}\right)\right.
\]

\begin{equation}
\left.-mR_{r}\Gamma\left(\frac{1}{2}+k{}_{ed_{3}}\right)\mathcal{H}\left(\left\{ \frac{1}{2}+k{}_{ed_{3}},\frac{1+m}{2},\frac{2+m}{2}\right\} ,\left\{ \frac{3}{2},\frac{3}{2}+K_{ed}\right\} ,R_{r}^{2}\right)\right)
\end{equation}

\[
\mathcal{E}\left\{ l_{r,u3u}^{-m}\right\} =\frac{\sqrt{\pi}\Gamma\left(1-k{}_{eu_{1}}+K_{eu}\right)}{4}\left(2\Gamma\left(k{}_{eu_{1}}\right)\mathcal{H}\left(\left\{ k{}_{eu_{1}},\frac{1+m}{2},\frac{m}{2}\right\} ,\left\{ \frac{1}{2},1+K_{eu}\right\} ,R_{r}^{2}\right)\right.
\]

\begin{equation}
-mR_{r}\Gamma\left(\frac{1}{2}+k{}_{eu_{1}}\right)\mathcal{H}\left(\left\{ \frac{1}{2}+k{}_{eu_{1}},\frac{1+m}{2},\frac{2+m}{2}\right\} ,\left\{ \frac{3}{2},\frac{3}{2}+K_{eu}\right\} ,R_{r}^{2}\right)
\end{equation}

and

\[
\mathscr{E}\left\{ \left|g_{u_{3u,r}}\Theta_{k}g_{_{r,u3d}}\right|^{2}\right\} =\frac{\kappa_{r,u3d}}{\kappa_{r,u3d}+1}\frac{\kappa_{u_{3u,r}}}{\kappa_{u_{3u,r}}+1}\xi_{6}+\frac{\kappa_{r,u3d}}{\kappa_{r,u3d}+1}\frac{1}{\kappa_{u_{3u,r}}+1}\stackrel[n=1]{N}{\sum}\left|\rho_{n}^{k}\right|^{2}
\]

\begin{equation}
+\frac{\kappa_{u_{3u,r}}}{\kappa_{u_{3u,r}}+1}\frac{1}{\kappa_{u_{3u,r}}+1}\stackrel[n=1]{N}{\sum}\left|\rho_{n}^{k}\right|^{2}+\frac{1}{\kappa_{r,u3d}+1}\frac{1}{\kappa_{u_{3u,r}}+1}\stackrel[n=1]{N}{\sum}\left|\rho_{n}^{k}\right|^{2}
\end{equation}

\noindent where $\xi_{6}=\left|\mathbf{\bar{g}}_{_{r,u3d}}\Theta\bar{\mathbf{g}}_{u_{3u,r}}\right|^{2}.$

\section*{Appendix D}

By using Jensen inequality, the ergodic rate can be approximated by

\[
\mathcal{E}\left[R_{u_{1u}}\right]\thickapprox\log_{2}\left(1+\right.
\]

\begin{equation}
\left.\frac{p_{u_{1u}}\mathcal{E}\left\{ A_{u_{1u}}\right\} }{p_{u_{2u}}\mathcal{E}\left\{ B_{u_{1u}}\right\} +p_{u_{3u}}\mathcal{E}\left\{ C_{u_{1u}}\right\} +P_{b}\mathcal{E}\left\{ D_{u_{1u}}\right\} +V+\sigma_{u_{1u}}^{2}}\right)
\end{equation}

1-The first term, $\mathcal{E}\left\{ A_{u_{1u}}\right\} $ can be
calculated as

\begin{equation}
\mathcal{E}\left\{ l_{b,u_{1u}}^{-m}\left|h_{b,u_{1u}}\right|^{2}\right\} =\mathcal{E}\left\{ l_{b,u_{1u}}^{-m}\right\} 
\end{equation}

which can be found as 

\begin{equation}
\mathcal{E}\left\{ l_{b,u_{1u}}^{-m}\right\} =\frac{2K_{cu}!}{\left(k_{cu_{1}}-1\right)!\left(K_{cu}-1\right)!}\stackrel[0]{R}{\int}r_{u_{1u}}^{-m}\frac{r}{\left(R\right)^{2}}\left(\frac{r^{2}}{\left(R\right)^{2}}\right)^{k-1}\left(1-\frac{r^{2}}{R^{2}}\right)^{K_{cu}-k_{cu_{1}}}dr_{u_{1u}}
\end{equation}

\[
\mathcal{E}\left\{ l_{b,u_{1u}}^{-m}\right\} =\frac{\sqrt{\pi}\Gamma\left(1-k_{cu_{1}}+K_{cu}\right)2K_{cu}!}{4\left(k_{cu_{1}}-1\right)!\left(K_{cu}-k_{cu_{1}}\right)!}\left(2\Gamma\left(k_{cu_{1}}\right)\mathcal{H}\left(\left\{ k_{cu_{1}},\frac{1+m}{2},\frac{m}{2}\right\} ,\left\{ \frac{1}{2},1+K_{cu}\right\} ,R^{2}\right)\right.
\]

\begin{equation}
\left.-mR\Gamma\left(\frac{1}{2}+k_{cu_{1}}\right)\mathcal{H}\left(\left\{ \frac{1}{2}+k_{cu_{1}},\frac{1+m}{2},\frac{2+m}{2}\right\} ,\left\{ \frac{3}{2},\frac{3}{2}+K_{cu}\right\} ,R^{2}\right)\right)\label{eq:8-3}
\end{equation}

2-The term, $\mathcal{E}\left\{ B_{u_{1u}}\right\} $ can be calculated
as

\[
\mathcal{E}\left\{ p_{u_{2u}}l_{b,u_{2u}}^{-m}\left|h_{b,u_{2u}}\right|^{2}\right\} =p_{u_{2u}}\mathcal{E}\left\{ l_{b,u_{2u}}^{-m}\right\} 
\]

\begin{equation}
\mathcal{E}\left\{ l_{b,u_{2u}}^{-m}\right\} =\frac{2K_{cu}!}{\left(k_{cu_{2}}-1\right)!\left(K_{cu}-1\right)!}\stackrel[0]{R}{\int}\left(r_{u_{2u}}\right)^{-m}\frac{\left(r\right)}{\left(R\right)^{2}}\left(\frac{r^{2}}{R^{2}}\right)^{k_{cu_{2}}-1}\left(1-\frac{r^{2}}{R^{2}}\right)^{K_{cu}-k_{cu_{2}}}dr_{u_{2u}}
\end{equation}

\[
\mathcal{E}\left\{ l_{b,u_{2u}}^{-m}\right\} =\frac{\sqrt{\pi}\Gamma\left(1-k_{cu_{2}}+K_{cu}\right)2K_{cu}!}{4\left(k_{cu_{2}}-1\right)!\left(K_{cu}-k_{cu_{2}}\right)!}\left(2\Gamma\left(k_{cu_{2}}\right)\mathcal{H}\left(\left\{ k_{cu_{2}},\frac{1+m}{2},\frac{m}{2}\right\} ,\left\{ \frac{1}{2},1+K_{cu}\right\} ,R^{2}\right)\right.
\]

\begin{equation}
\left.-mR\Gamma\left(\frac{1}{2}+k_{cu_{2}}\right)\mathcal{H}\left(\left\{ \frac{1}{2}+k_{cu_{2}},\frac{1+m}{2},\frac{2+m}{2}\right\} ,\left\{ \frac{3}{2},\frac{3}{2}+K_{cu}\right\} ,R^{2}\right)\right)\label{eq:8-2-1}
\end{equation}

3- Similarly, the term, $\mathcal{E}\left\{ C_{u_{1u}}\right\} $
can be calculated as

\begin{equation}
\mathcal{E}\left\{ l_{b,r}^{-m}l_{r,u3u}^{-m}\left|\mathbf{g}_{b,r}\Theta\mathbf{g}_{_{r,u3u}}\right|^{2}\right\} =l_{b,r}^{-m}\mathcal{E}\left\{ l_{r,u3u}^{-m}\right\} \mathcal{E}\left\{ \left|\mathbf{g}_{b,r}\Theta\mathbf{g}_{_{r,u3u}}\right|^{2}\right\} 
\end{equation}

where

\[
\mathcal{E}\left\{ l_{r,u3u}^{-m}\right\} =\frac{\sqrt{\pi}\Gamma\left(1-k_{eu_{1}}+K_{eu}\right)}{4}\left(2\Gamma\left(k_{eu_{1}}\right)\mathcal{H}\left(\left\{ k_{eu_{1}},\frac{1+m}{2},\frac{m}{2}\right\} ,\left\{ \frac{1}{2},1+K_{eu}\right\} ,R_{r}^{2}\right)\right.
\]

\begin{equation}
\left.-mR_{r}\Gamma\left(\frac{1}{2}+k_{eu_{1}}\right)\mathcal{H}\left(\left\{ \frac{1}{2}+k_{eu_{1}},\frac{1+m}{2},\frac{2+m}{2}\right\} ,\left\{ \frac{3}{2},\frac{3}{2}+K_{eu}\right\} ,R_{r}^{2}\right)\right)
\end{equation}

\[
\mathscr{E}\left\{ \left|\mathbf{g}_{b,r}\Theta\mathbf{g}_{_{r,u3u}}\right|^{2}\right\} =\frac{\kappa_{r,u3u}}{\kappa_{r,u3u}+1}\frac{\kappa_{b,r}}{\kappa_{b,r}+1}\xi_{7}+\frac{\kappa_{r,u3u}}{\kappa_{r,u3u}+1}\frac{1}{\kappa_{b,r}+1}\stackrel[n=1]{N}{\sum}\left|\rho_{n}^{k}\right|^{2}
\]

\begin{equation}
+\frac{\kappa_{b,r}}{\kappa_{b,r}+1}\frac{1}{\kappa_{r,u3u}+1}\stackrel[n=1]{N}{\sum}\left|\rho_{n}^{k}\right|^{2}+\frac{1}{\kappa_{r,u3u}+1}\frac{1}{\kappa_{b,r}+1}\stackrel[n=1]{N}{\sum}\left|\rho_{n}^{k}\right|^{2}
\end{equation}

\noindent where $\xi_{7}=\left|\bar{\mathbf{g}}_{b,r}\Theta\mathbf{\bar{g}}_{_{_{r,u3u}}}\right|^{2}.$

4- The term, $\mathcal{E}\left\{ D_{u_{1u}}\right\} $ can be calculated
as

\[
\mathcal{E}\left\{ l_{b,r}^{-m}l_{b,r}^{-m}\left|\mathbf{g}_{b,r}\Theta_{k}\mathbf{g}_{b,r}^{H}\right|^{2}\right\} =l_{b,r}^{-m}l_{b,r}^{-m}\mathcal{E}\left\{ \left|\mathbf{g}_{b,r}\Theta_{k}\mathbf{g}_{b,r}^{H}\right|^{2}\right\} 
\]

\[
\mathscr{E}\left\{ \left|\mathbf{g}_{b,r}\Theta_{k}\mathbf{g}_{b,r}^{H}\right|^{2}\right\} =\mathscr{E}\left\{ \left|\left(\sqrt{\frac{\kappa_{b,r}}{\kappa_{b,r}+1}}\sqrt{\frac{\kappa_{b,r}}{\kappa_{b,r}+1}}\mathbf{\bar{g}}_{b,r}\Theta\bar{\mathbf{g}}_{b,r}^{H}+\sqrt{\frac{\kappa_{b,r}}{\kappa_{b,r}+1}}\sqrt{\frac{1}{\kappa_{b,r}+1}}\mathbf{\bar{g}}_{b,r}\Theta\mathbf{\tilde{g}}_{b,r}^{H}\right.\right.\right.
\]

\begin{equation}
\left.\left.\left.+\sqrt{\frac{\kappa_{b,r}}{\kappa_{b,r}+1}}\sqrt{\frac{1}{\kappa_{b,r}+1}}\mathbf{\tilde{g}}_{b,r}\Theta\bar{\mathbf{g}}_{b,r}^{H}+\sqrt{\frac{1}{\kappa_{b,r}+1}}\sqrt{\frac{1}{\kappa_{b,r}+1}}\mathbf{\tilde{g}}_{b,r}\Theta\mathbf{\tilde{g}}_{b,r}^{H}\right)\right|^{2}\right\} 
\end{equation}

\[
\mathscr{E}\left\{ \left|\mathbf{g}_{b,r}\Theta_{k}\mathbf{g}_{b,r}^{H}\right|^{2}\right\} =\left(\frac{\kappa_{b,r}}{\kappa_{b,r}+1}\right)^{2}\mathscr{E}\left|\mathbf{\bar{g}}_{b,r}\Theta\bar{\mathbf{g}}_{b,r}^{H}\right|^{2}+\frac{\kappa_{b,r}}{\kappa_{b,r}+1}\frac{1}{\kappa_{b,r}+1}\mathscr{E}\left|\mathbf{\bar{g}}_{b,r}\Theta\mathbf{\tilde{g}}_{b,r}^{H}\right|^{2}
\]

\[
+\frac{\kappa_{b,r}}{\kappa_{b,r}+1}\frac{1}{\kappa_{b,r}+1}\mathscr{E}\left|\mathbf{\tilde{g}}_{b,r}\Theta\bar{\mathbf{g}}_{b,r}^{H}\right|^{2}+\left(\frac{1}{\kappa_{b,r}+1}\right)^{2}\mathscr{E}\left|\mathbf{\tilde{g}}_{b,r}\Theta\mathbf{\tilde{g}}_{b,r}^{H}\right|^{2}
\]

\begin{equation}
+2\sqrt{\frac{\kappa_{b,r}}{\kappa_{b,r}+1}}\sqrt{\frac{\kappa_{b,r}}{\kappa_{b,r}+1}}\mathscr{E}\left(\mathbf{\bar{g}}_{b,r}\Theta\bar{\mathbf{g}}_{b,r}^{H}\left(\mathbf{\tilde{g}}_{b,r}\Theta\mathbf{\tilde{g}}_{b,r}^{H}\right)^{H}\right)
\end{equation}

Now

\begin{equation}
\mathscr{E}\left|\mathbf{\bar{g}}_{b,r}\Theta\bar{\mathbf{g}}_{b,r}^{H}\right|^{2}=\left|\mathbf{\bar{g}}_{b,r}\Theta\bar{\mathbf{g}}_{b,r}^{H}\right|^{2}=\xi_{8}
\end{equation}

Similarly, the second term,

\begin{equation}
\mathbf{\bar{g}}_{b,r}\Theta\mathbf{\tilde{g}}_{b,r}^{H}=\stackrel[n=1]{N}{\sum}a_{Nn}\left(\psi_{b,r}^{a},\psi_{b,r}^{e}\right)\rho_{n}^{k}e^{j\phi_{n}^{k}}\left[\tilde{\mathbf{g}}_{b,r}\right]_{n}
\end{equation}

\begin{equation}
\mathscr{E}\left|\mathbf{\bar{g}}_{b,r}\Theta\mathbf{\tilde{g}}_{b,r}^{H}\right|^{2}=\mathscr{E}\left|\stackrel[n=1]{N}{\sum}a_{Nn}\left(\psi_{b,r}^{a},\psi_{b,r}^{e}\right)\rho_{n}^{k}e^{j\phi_{n}^{k}}\left[\tilde{\mathbf{g}}_{b,r}\right]_{n}\right|^{2}
\end{equation}

\[
\mathscr{E}\left|\mathbf{\bar{g}}_{b,r}\Theta\mathbf{\tilde{g}}_{b,r}^{H}\right|^{2}=\stackrel[n=1]{N}{\sum}\left|\rho_{n}^{k}\right|^{2}+
\]

\begin{equation}
\mathscr{E}\left\{ \stackrel[n_{1}=1]{N}{\sum}\stackrel[n_{2}\neq n_{1}]{N}{\sum}\left(a_{Nn_{1}}\left(\psi_{b,r}^{a},\psi_{b,r}^{e}\right)\rho_{n_{1}}^{k}e^{j\phi_{n_{1}}^{k}}\left[\tilde{\mathbf{g}}_{b,r}\right]_{n_{1}}\right)\left(a_{Nn_{2}}\left(\psi_{b,r}^{a},\psi_{b,r}^{e}\right)\rho_{n_{2}}^{k}e^{j\phi_{n_{2}}^{k}}\left[\tilde{\mathbf{g}}_{b,r}\right]_{n_{2}}\right)^{H}\right\} =\stackrel[n=1]{N}{\sum}\left|\rho_{n}^{k}\right|^{2}
\end{equation}

Similarly,

\begin{equation}
\mathbf{\tilde{g}}_{b,r}\Theta\bar{\mathbf{g}}_{b,r}^{H}=\stackrel[n=1]{N}{\sum}\left[\mathbf{\tilde{g}}_{b,r}\right]_{n}\rho_{n}^{k}e^{j\phi_{n}^{k}}a_{N,n}\left(\psi_{b,r}^{a},\psi_{b,r}^{e}\right)
\end{equation}

\[
\mathscr{E}\left|\mathbf{\tilde{g}}_{b,r}\Theta\bar{\mathbf{g}}_{b,r}^{H}\right|^{2}=\stackrel[n=1]{N}{\sum}\left|\rho_{n}^{k}\right|^{2}+
\]

\begin{equation}
\mathscr{E}\left\{ \stackrel[n_{1}=1]{N}{\sum}\stackrel[n_{2}\neq n_{1}]{N}{\sum}\left(\left[\mathbf{\tilde{g}}_{b,r}\right]_{n_{1}}\rho_{n_{1}}^{k}e^{j\phi_{n_{1}}^{k}}a_{Nn1}\left(\psi_{b,r}^{a},\psi_{b,r}^{e}\right)\right)\left(\left[\mathbf{\tilde{g}}_{b,r}\right]_{n_{2}}\rho_{n_{1}}^{k}e^{j\phi_{n_{1}}^{k}}a_{Nn_{2}}\left(\psi_{b,r}^{a},\psi_{b,r}^{e}\right)\right)^{H}\right\} =\stackrel[n=1]{N}{\sum}\left|\rho_{n}^{k}\right|^{2}
\end{equation}

and

\begin{equation}
\mathbf{\tilde{g}}_{b,r}\Theta\mathbf{\tilde{g}}_{b,r}^{H}=\stackrel[n=1]{N}{\sum}\left[\mathbf{\tilde{g}}_{b,r}\right]_{n}\rho_{n}^{k}e^{j\phi_{n}^{k}}\left[\mathbf{\tilde{g}}_{b,r}^{H}\right]_{n}
\end{equation}

\begin{equation}
\mathscr{E}\left|\mathbf{\tilde{g}}_{b,r}\Theta\mathbf{\tilde{g}}_{b,r}^{H}\right|^{2}=\mathscr{E}\left|\stackrel[n=1]{N}{\sum}\left[\mathbf{\tilde{g}}_{b,r}\right]_{n}\rho_{n}^{k}e^{j\phi_{n}^{k}}\left[\mathbf{\tilde{g}}_{b,r}^{H}\right]_{n}\right|^{2}
\end{equation}

\[
\mathscr{E}\left|\mathbf{\tilde{g}}_{b,r}\Theta\mathbf{\tilde{g}}_{b,r}^{H}\right|^{2}=2\stackrel[n=1]{N}{\sum}\left|\rho_{n}^{k}\right|^{2}+
\]

\[
\mathscr{E}\left\{ \stackrel[n_{1}=1]{N}{\sum}\stackrel[n_{2}\neq n_{1}]{N}{\sum}\left(\left[\tilde{\mathbf{g}}_{b,r}\right]_{n_{1}}\rho_{n_{1}}^{k}e^{j\phi_{n_{1}}^{k}}\left[\tilde{\mathbf{g}}_{b,r}^{H}\right]_{n_{1}}\right)\left(\left[\tilde{\mathbf{g}}_{b,r}\right]_{n_{2}}\rho_{n_{2}}^{k}e^{j\phi_{n_{2}}^{k}}\left[\tilde{\mathbf{g}}_{b,r}^{H}\right]_{n_{2}}\right)^{H}\right\} =2\stackrel[n=1]{N}{\sum}\left|\rho_{n}^{k}\right|^{2}
\]

\begin{equation}
+\stackrel[n_{1}=1]{N}{\sum}\stackrel[n_{2}\neq n_{1}]{N}{\sum}\left(\rho_{n_{1}}^{k}e^{j\phi_{n_{1}}^{k}}\right)\left(\rho_{n_{2}}^{k}e^{j\phi_{n_{2}}^{k}}\right)^{H}
\end{equation}

Last term

\[
\mathscr{E}\left(\mathbf{\bar{g}}_{b,r}\Theta\bar{\mathbf{g}}_{b,r}^{H}\left(\mathbf{\tilde{g}}_{b,r}\Theta\mathbf{\tilde{g}}_{b,r}^{H}\right)^{H}\right)=\stackrel[n=1]{N}{\sum}a_{Nn}\left(\psi_{b,r}^{a},\psi_{b,r}^{e}\right)^{H}\rho_{n}^{k}e^{j\phi_{n}^{k}}a_{Nn}\left(\psi_{b,r}^{a},\psi_{b,r}^{e}\right)\stackrel[n=1]{N}{\sum}\left(\rho_{n}^{k}e^{j\phi_{n}^{k}}\right)^{H}
\]

\begin{equation}
=\mathbf{\bar{g}}_{b,r}\Theta\bar{\mathbf{g}}_{b,r}^{H}\stackrel[n=1]{N}{\sum}\rho_{n}^{k}e^{j\phi_{n}^{k}}=\hat{\zeta}\stackrel[n=1]{N}{\sum}\left(\rho_{n}^{k}e^{j\phi_{n}^{k}}\right)^{H}
\end{equation}

\bibliographystyle{IEEEtran}
\bibliography{J_One_Col}

\end{document}